\documentclass[letterpaper,showpacs, onecolumn,eqsecnum,superscriptaddress,floatfix]{revtex4}
\pdfoutput=1

\usepackage{amssymb}
\usepackage{amsmath}
\usepackage{latexsym}
\usepackage{epsfig}
\usepackage{color}
\usepackage{graphicx,subfigure}
\usepackage{units}
\usepackage{natbib}
\usepackage[linktocpage]{hyperref}

\begin{document}

%%%%%%%%%%%%%%%%%
%%%   TITLE   %%%
%%%%%%%%%%%%%%%%%
\title{Gravitational and electromagnetic signatures of accretion into a 
charged black hole}

\author{Claudia Moreno} 
\email[]{claudia.moreno@cucei.udg.mx} 
\affiliation{Departamento de F\'isica,
Centro Universitario de Ciencias Exactas e Ingenier\'ias, Universidad de Guadalajara\\
Av. Revoluci\'on 1500, Colonia Ol\'impica C.P. 44430, Guadalajara, Jalisco, M\'exico. }
\affiliation{Instituto de F\'isica y Matem\'aticas,
Universidad Michoacana de San Nicol\'as de Hidalgo,\\
Edificio C-3, Ciudad Universitaria, 58040 Morelia, Michoac\'an, M\'exico.}

\author{Juan Carlos Degollado} 
\email[]{jcdegollado@ciencias.unam.mx}
\affiliation{
Instituto de Ciencias F\'isicas, Universidad Nacional Aut\'onoma de M\'exico,
Apartado Postal 48-3, 62251, Cuernavaca, Morelos, M\'exico.}

\author{Dar\'{\i}o N\'u\~nez}
\email[]{nunez@nucleares.unam.mx}
\affiliation{Instituto de Ciencias Nucleares, Universidad Nacional
  Aut\'onoma de M\'exico, Circuito Exterior C.U., Apartado Postal 70-543,
  Ciudad de M\'exico, 04510, M\'exico.}

%%%%%%%%%%%%%%%%
%%%   DATE   %%%
%%%%%%%%%%%%%%%%

\date{\today}

%%%%%%%%%%%%%%%%%%%%
%%%   ABSTRACT   %%%
%%%%%%%%%%%%%%%%%%%%

\bigskip

\begin{abstract}
We present the derivation and the solutions to the coupled electromagnetic and 
gravitational perturbations
with sources in a charged black hole background. We work in the so called {\it ghost gauge} and
consider as source of the perturbations the infall of radial currents. In this 
way, we study a system in which it
is provoked a response involving both, gravitational and electromagnetic 
waves, which
allows us to analyze the dependence between them. 
We solve numerically the wave equations that describe both signals, 
characterize the waveforms and study the relation between the input 
parameters of the infalling matter with those of the gravitational and 
electromagnetic responses.
\end{abstract}

%%%%%%%%%%%%%%%%
%%%   PACS   %%%
%%%%%%%%%%%%%%%%


%%%%%%%%%%%%%%%%%%%%%%
%%%   MAKE TITLE   %%%
%%%%%%%%%%%%%%%%%%%%%%

\maketitle

%%%%%%%%%%%%%%%%%%%%%%%%
%%%   INTRODUCTION   %%%
%%%%%%%%%%%%%%%%%%%%%%%%

\section{Introduction}
\label{sec:introduction}

Gravitational waves emitted by distorted black holes carry 
information about the corresponding space-time. In some cases this signal may 
have an electromagnetic counterpart and one can either use electromagnetic or
gravitational wave observations to probe strong-field 
gravity with Black Holes~\cite{Dimitrios-2008,Bambi:2015kza,Kamble2016}. 
Regarding the 
former, one can use X-ray continuum spectra and Fe K$\alpha$ line emissions to 
infer black hole spins also, such spectra can be used to probe deviations from 
the black hole solutions of 
General Relativity since the spectra depends on the position of the inner edge 
of the accretion disk around the black hole. Gravitational waves on the other hand may 
be used to follow the dynamics of the collision of two black holes as was done in the 
recent detection GW150914 and GW151226 by the LIGO-Virgo collaboration 
\cite{Abbott:2016blz,TheLIGOScientific:2016qqj,Abbott:2016nmj}.

There are scenarios in which is expected an 
electromagnetic counterpart associated with a large emission of gravitational 
wave, for instance the short duration gamma ray burst related 
with the double black hole merger GW150914 as described in 
\cite{Connaughton:2016umz}. 
Electromagnetic signals associated with gravitational wave events are thus an 
interesting case of study. The interaction between electromagnetic radiation 
and gravitational waves gave rise to multimessenger astrophysics. An extensive 
study involving perturbations of neutron stars and black holes have been 
done in \cite{Diaz-2016,Sotani:2013iha,Sotani:2014fia}.

In General Relativity, black holes are uniquely described by three parameters, 
mass $M$, angular momentum $J$, and charge $Q$ \cite{Wald84}. Whereas the first 
two parameters 
have been estimated with various observations, it has been widely believed that 
the charge must be very small or null. Due to the interaction with the 
interstellar media, a charged black hole should quickly discharge. Nevertheless 
charged spherical black holes allow us to set up scenarios in which 
electromagnetic waves are produced in conjunction with gravitational waves 
such as the collision of binary charged black 
holes~\cite{Alcubierre:2009ij,Zilhao:2012gp,Zilhao:2013nda}.

One of the preferred models of charged black holes is  
the Reissner-N\"ordstrom  spacetime, which is a stationary spherically 
symmetric 
solution of the coupled Einstein-Maxwell system. 
It represents a black hole when $Q<M$ and it has been shown that  
is stable under linear perturbations~\cite{Gunter497, Moncrief:1974ng}. 
The importance of this spacetime relies in the fact that, when it is perturbed, 
one may expect a significant amount of energy released in form of gravitational 
and electromagnetic waves.

There are several studies involving perturbations of Reissner-N\"ordstrom black 
holes. Some of them have 
been made considering perturbations of the components of the metric, as in
the pioneering work of Zerilli and Moncrief 
\cite{Zerilli:1974ai,Moncrief:1974gw,Moncrief75}, or 
considering perturbations of  the scalars of curvature given by projections of 
the Weyl tensor. The later is know as the Newman Penrose 
formalism~\cite{Newman62a,Chandrasekhar83, Chandrasekhar79} and was used by 
Teukolsky to derive a master perturbation equation for rotating black 
holes~\cite{Teukolsky:1973ha, Teukolsky:1974yv} including electromagnetic 
radiation (see also~\cite{Press73,Kodama:2003kk} and references therein).

One of the most remarkable results of perturbation theory applied to the  
Reissner-N\"ordstrom solution is that the gravitational 
and electromagnetic perturbations can not be decoupled \cite{LeeCH1977, 
LeeCH2012}. By virtue of this coupling the energy of an incident purely 
gravitational wave, is 
reflected in part as electromagnetic waves and conversely. This transformation
of gravitational energy into electromagnetic energy is encoded in a  
coupled set of equations for both perturbations.

In this paper we find the expressions for the perturbations of the 
Reissner-N\"ordstrom 
black hole in the Newman Penrose formalism including the matter 
sources that may cause the perturbations. We use the expansion of the 
perturbed fields in terms of spherical harmonics. We focus on a model in which  
the source of 
perturbation is a set of 
charged particles falling into the black hole. Then, we solve numerically the equations 
of gravitation-electromagnetic perturbations coupled to the equations of the 
dynamics of the particles. We study the dependence of the coupled 
electromagnetic and gravitational waveforms on the parameters of the accreting 
fluid, such as its charge and density profile.

The paper is organized as follows: In section \ref{sec:foundations} we 
introduce the concepts of the null tetrad formalism. Next, in section 
\ref{sec:PME} we derive the perturbed Maxwell equations 
and in the following section \ref{sec:PBI} we explicitly derive the outgoing 
coupled 
gravitational-electromagnetic equations. In \ref{sec:CES_RN} we introduce 
the tetrad and geometric quantities of 
relevance in the Reissner-N\"ordstrom background described in horizon 
penetrating coordinates, and we write the coupled system of  
equations for the perturbed Weyl scalars ${\Psi_4}^{(1)}, {\Psi_3}^{(1)}$ with 
sources. In \ref{sec:MD} we describe the dynamics of the matter that cause the 
perturbations. 
In sections \ref{sec:numerical} and \ref{sec:results}
we introduce the numerical procedure to solve 
the equations and analyze the properties of the waveforms.
Finally, in section \ref{sec:conclusions} we give some concluding 
remarks. 

In this work we adopt geometric units $c = G =4\pi\epsilon_0= 1 $ and the 
metric signature $(-, +, +, +)$. In the derivations, however, we will include 
the signature $(+,-,-,-)$ in order to be able to compare with the results 
obtained in classical works such as \cite{Chandrasekhar83}.

%%%%%%%%%%%%%%%%%%%%%%%%
%%%   FOUNDATIONS   %%%
%%%%%%%%%%%%%%%%%%%%%%%%
\section{Foundations: Newman-Penrose formalism}
\label{sec:foundations}

The Newman Penrose formalism is particularly suited for dealing with radiation 
in asymptotically flat space-times since ingoing and outgoing radiation are well 
represented by the Weyl scalars.
We introduce a Newman-Penrose null tetrad $l^{\mu},k^{\mu}, m^{\mu}$, 
choosing it in such way that $l^\mu$ is pointing outward and $k^\mu$ is 
pointing inward in the asymptotic 
region of a hyper-surface of constant 
time. The directional operators are defined as in Chandrasekhar's book 
\cite{Chandrasekhar83}:
\begin{equation}
 D = l^{\mu}\partial_{\mu}, \qquad \Delta = k^{\mu}\partial_{\mu},\qquad 
\delta = m^{\mu}\partial_{\mu}, \qquad {\rm and}\qquad  \overline\delta = 
\overline m^{\mu}\partial_{\mu},
\end{equation}
as well as the the spinor coefficients:
\begin{eqnarray}
\kappa_s&=&\gamma_{311}=m^\mu\,l_{\mu;\nu}\,l^\nu; \hspace{0.5cm} \tau_s=\gamma_{312}=m^\mu\,l_{\mu;\nu}\,k^\nu; 
\hspace{0.5cm} \sigma_s=\gamma_{313}=m^\mu\,l_{\mu;\nu}\,m^\nu; \hspace{0.5cm} 
\rho_s=\gamma_{314}=m^\mu\,l_{\mu;\nu}\,{\overline m}^\nu; \nonumber \\
\pi_s&=&\gamma_{241}=k^\mu\,{\overline m}_{\mu;\nu}\,l^\nu; \hspace{0.5cm} 
\nu_s=\gamma_{242}=k^\mu\,{\overline m}_{\mu;\nu}\,k^\nu; 
\hspace{0.5cm} \mu_s=\gamma_{243}=k^\mu\,{\overline m}_{\mu;\nu}\,m^\nu; 
\hspace{0.5cm} \lambda_s=\gamma_{244}=k^\mu\,{\overline 
m}_{\mu;\nu}\,{\overline m}^\nu; \nonumber \\
\epsilon_s&=&\frac12\left(\gamma_{211} + 
\gamma_{341}\right)=\frac12\left(k^\mu\,l_{\mu;\nu} + m^\mu\,{\overline 
m}_{\mu;\nu}\right)\,l^\nu; 
\hspace{0.5cm} \gamma_s=\frac12\left(\gamma_{212} + 
\gamma_{342}\right)=\frac12\left(k^\mu\,l_{\mu;\nu} + m^\mu\,{\overline 
m}_{\mu;\nu}\right)\,k^\nu;\nonumber \\
\beta_s&=&\frac12\left(\gamma_{213} + 
\gamma_{343}\right)=\frac12\left(k^\mu\,l_{\mu;\nu} + m^\mu\,{\overline 
m}_{\mu;\nu}\right)\,m^\nu; 
\hspace{0.5cm} \alpha_s=\frac12\left(\gamma_{214} + 
\gamma_{344}\right)=\frac12\left(k^\mu\,l_{\mu;\nu} + 
m^\mu\,{\overline m}_{\mu;\nu}\right)\,{\overline m}^\nu. 
\label{eq:coef_spinors}
\end{eqnarray}

The curvature quantities $\Psi$'s and $\Phi$'s and the electromagnetic 
scalars, $\varphi$,  are defined as 
\begin{eqnarray}
\Psi_4&=&-C_{\mu\nu\lambda\tau}\,k^\mu\,\overline{m}^\nu\,k^\lambda\,\overline{m}^\tau; 
\hspace{0.5cm} \Psi_3=-C_{\mu\nu\lambda\tau}\,l^\mu\,k^\nu\,\overline{m}^\lambda\,{k}^\tau; \hspace{0.5cm} 
\Psi_2=-C_{\mu\nu\lambda\tau}\,l^\mu\,{m}^\nu\,\overline{m}^\lambda\,{k}^\tau\\
\Phi_{00} &=&\overline \Phi_{00} =\frac{1}{2}R_{\mu \nu} l^{\mu}l^{\nu}= 4 \pi T_{\mu \nu} l^\mu l^\nu \equiv 4\pi T_{ll}, \hspace{0.5cm} 
\Phi_{01} = {\overline \Phi_{10}}=\frac{1}{2}R_{\mu \nu} l^{\mu}m^{\nu}= 4 \pi T_{\mu \nu} l^\mu m^\nu \equiv 4\pi T_{lm},   \nonumber \\
\Phi_{02} &=& {\overline \Phi_{20}} =\frac{1}{2}R_{\mu \nu} m^{\mu}m^{\nu}= 4 
\pi T_{\mu \nu} m^\mu m^\nu \equiv 4\pi T_{mm}, \nonumber \\
\Phi_{22} &=&\overline \Phi_{22} = \frac{1}{2}R_{\mu \nu} n^{\mu}n^{\nu}= 4 \pi 
T_{\mu \nu} n^\mu n^\nu \equiv 4\pi T_{nn}, \label{ricci}\\
\varphi_0&=&F_{\mu\nu}l^{\mu}m^{\mu}; \hspace{0.5cm} \varphi_1:=\frac{1}{2} F_{\mu\nu}(l^{\mu}n^{\mu}+
\overline m^{\mu} m^{\nu}); \hspace{0.5cm} \varphi_2:=F_{\mu\nu}\overline m^{\mu}k^{\nu}. \label{proy_faraday}
\end{eqnarray}
with $C_{\mu\nu\lambda\tau}$ the Weyl tensor, $F_{\mu\nu}$ the Faraday tensor 
and $T_{\mu\nu}$ is the stress energy tensor of the matter content. 
These definitions are given in Newman and Penrose paper 
\cite{Newman62a}.
In general, the null tetrad vectors ${Z_a}^\mu$ satisfy 
the following normalization equations
\begin{equation}
{Z_a}^\mu\,{Z_b}_\mu=\eta_{ab}, \hspace{1cm} g_{\mu\nu}=2\eta^{ab}\,Z_{a\,(\mu}\,Z_{b\,\nu)},\label{eqs:tetrad}
\end{equation}
where $\eta_{ab}$ has the form:
\begin{equation}
\eta_{ab}=\left(\begin{matrix}&0&\, &\eta&\,&0&\,&0&\\&\eta&\,&0&\,&0&\,&0&\\&0&\,&0&\,&0&\,&-
\eta&\\&0&\,&0&\,&-\eta&\,&0&\end{matrix}\right),
\end{equation}
with $\eta$ a constant related to the signature. 
For the signature $(+,-,-,-)$, $\eta=1$ and for the signature 
$(-,+,+,+)$, $\eta=-1$. In this work, we derive the equations for the 
electromagnetic and for the gravitational 
perturbations considering both signatures, displaying $\eta$ explicitly in all 
the 
derivations, until we specify the components of the metric, from where we set 
$\eta=-1$.

With this set-up, what follows is to project the Maxwell equations, 
the Riemann, and the Weyl curvature tensors, the Bianchi identities and the  
Einstein equations on the tetrad to obtain the Newman-Penrose 
equations. Such procedure is described in detail for instance in 
\cite{Chandrasekhar83,Degollado:2011gi}.

%%%%%%%%%%%%%%%%%%%%%%%%%%%%%%%%%%%%%%%%%%%%%%%%%%%%%%%%%%%%%%%%%%%%%%
\section{Perturbed Maxwell Equations}
\label{sec:PME}
%%%%%%%%%%%%%%%%%%%%%%%%%%%%%%%%%%%%%%%%%%%%%%%%%%%%%%%%%%%%%%%%%%%%%%

Let us consider first, the electromagnetic scalars, Eq.~(\ref{proy_faraday}). 
These scalars are related with the ingoing and outgoing electromagnetic 
radiation at infinity and its dynamics is dictated by the 
Maxwell equations with sources, which are: 
\begin{eqnarray}
(D-2 \eta \rho_s) \varphi_1 - (\overline{\delta} + \eta(\pi_s - 2 \alpha_s)) 
\varphi_{0}+\eta \kappa_s \varphi_2= 2\,\eta\,  \pi J_l, \label{Muno} \\
(\delta-2 \eta \tau_s) \varphi_1 - (\Delta+ \eta(\mu_s - 2 \gamma_s)) 
\varphi_{0}+\eta \sigma_s \varphi_2= 2\,\eta\, \pi J_m, \label{Mdos} \\
(D- \eta(\rho_s - 2 \epsilon_s)) \varphi_2 - (\overline{\delta} + 2 \eta \pi_s ) \varphi_{1}+\eta \lambda_s \varphi_0= 
2\,\eta\, \pi J_{\overline{m}}, \label{Mtres} \\
(\delta-\eta(\tau_s + 2 \beta_s)) \varphi_2 - ( \Delta + 2 \eta \mu_s) \varphi_{1}
+\eta \nu_s \varphi_0= 2 \,\eta\, \pi J_k, \label{Mcuatro}
\end{eqnarray}
where $J_l = J_{\mu} l^{\mu}, J_m = J_{\mu} m^{\mu}$ and $J_k = J_{\mu} 
k^{\mu}$. 
$J_{\mu}$ is the 4-electric current \cite{Newman62a}.

To obtain an equation to describe the electromagnetic perturbations, we operate 
with $(\Delta +\eta(\overline{\mu}_s-\overline{\gamma}_s+ \gamma_s + 2\mu_s))$ on 
Eq. (\ref{Mtres}) and with  $(\overline \delta-\eta(\overline{\tau}_s - \alpha_s 
- \overline{\beta}_s - 2 \pi_s))$ on Eq.  
(\ref{Mcuatro}). After subtracting one equation from the other we get
\begin{eqnarray}
&&[ (\Delta +\eta(\overline{\mu}_s-\overline{\gamma}_s +  \gamma_s + 2\mu_s))(D-\eta (\rho_s-2\epsilon_s))-
(\overline{\delta}-\eta(\overline{\tau}_s - \alpha_s 
- \overline{\beta}_s-2\pi_s))(\delta -\eta(\tau_s - 2\beta_s))] \varphi_2 + \nonumber \\
&&- [(\Delta +\eta(\overline{\mu}_s-\overline{\gamma}_s+ \gamma_s + 2\mu_s))(\overline{\delta}+2\eta \pi_s)-(\overline{\delta}-\eta(\overline{\tau}_s
-\alpha_s - \overline{\beta}_s-2\pi_s))(\Delta+2\eta \mu_s)]\varphi_1 + \nonumber  \\
&&[(\Delta +\eta(\overline{\mu}_s-\overline{\gamma}_s+  \gamma_s + 2\mu_s))\eta \lambda_s-(\overline{\delta}-\eta(\overline{\tau}_s-\alpha_s 
- \overline{\beta}_s-2\pi_s))\eta \nu_s]\varphi_0=2\,\eta\,\hat\pi J_2, \label{phi3}
\end{eqnarray}
where
\begin{eqnarray}
J_2&=&(\Delta+\eta(\gamma_s-\overline{\gamma}_s +2 \mu_s + \overline{\mu}_s))J_{\overline {m}}
-(\overline{\delta} +\eta(\alpha_s +\overline{\beta}_s + 2 \pi_s -\overline{\tau}_s)) J_n. 
\nonumber 
\end{eqnarray}%
Expanding the term acting on $\varphi_1$ in Eq.~(\ref{phi3}) we have
{\small
\begin{eqnarray}
&&[(\Delta + \eta(\overline{\mu}_s - \overline{\gamma}_s + \gamma_s + 
2\mu_s))(D-\eta(\rho_s-2\epsilon_s))-
(\overline{\delta} - \eta(\overline{\tau}_s -\alpha_s - \overline{\beta}_s- 2 
\pi_s))
(\delta -\eta(\tau_s -2\beta_s))]\varphi_2 - [[\Delta, \overline{\delta} 
]\varphi_1 + 2\eta(\Delta \pi_s) \varphi_1 \nonumber 
\\ &&- 2\eta(\overline{\delta} \mu_s) \varphi_1 + 2\eta \pi_s(\Delta \varphi_1)  
- 
2\eta \mu_s\,(\overline{\delta} \varphi_1) +
\eta(\gamma_s-\overline{\gamma}_s+2\mu_s + \overline{\mu}_s)(\overline{\delta} + 
2\eta\delta \pi_s) \varphi_1 - 
\eta (\alpha_s+ \overline{\beta}_s + 2\pi_s -\overline{\tau}_s) (\Delta +2 \eta 
\mu_s)\varphi_1 ] + \nonumber \\ 
&& \varphi_0\,[(\Delta + \eta (\overline{\mu}_s -\overline{\gamma}_s + \gamma_s 
+ 2\mu_s))\eta \lambda_s 
-(\overline{\delta} - \eta (\overline{\tau}_s - \alpha_s -\overline{\beta}_s - 
2\pi_s))\eta \nu_s]+\eta \lambda_s \Delta \varphi_0
-\eta\nu_s \overline{\delta} \varphi_0  = 2\eta\, \pi J_2. \label{eq:phi2}
\end{eqnarray}
}

As is a common practice~\cite{Chandrasekhar83}, we use the 
commutation properties of the differential operators, for instance,
replacing the commutator $[\Delta, \overline{\delta}]$,
\begin{equation}
 [\Delta, \overline{\delta} ] = \eta \nu_s D + \eta (\alpha_s + \overline{\beta}_s - 
 \overline{\tau}_s) \Delta - \eta (\overline{\mu}_s-\overline{\gamma}_s+ \gamma_s)\overline \delta - \eta \lambda_s \delta, 
 \label{eq:con_Dd}
\end{equation}
and the Ricci background identities $\Delta \pi_s,$ $\overline{\delta}\mu_s$, 
\begin{eqnarray}
\eta \Delta \pi_s &=& \eta D\nu_s -\mu_s(\pi_s+\overline{\tau}_s) - \lambda_s(\overline{\pi}_s+\tau_s) - 
\pi_s (\gamma_s-\overline{\gamma}_s)+\nu_s (3\epsilon_s+\overline{\epsilon})- \eta \Psi_3 - \Phi_{21}, \nonumber \\
\eta \overline{\delta} \mu_s&=&\eta \delta \lambda_s -\nu_s(\rho_s-\overline{\rho}_s)-\pi_s(\mu_s-\overline{\mu})-
\mu_s(\alpha_s+\overline{\beta}_s)
-\lambda_s(\overline{\alpha}_s-3\beta_s) + \eta \Psi_3-\Phi_{21},
\end{eqnarray}
with such substitutions, equation~(\ref{eq:phi2}) becomes
\begin{eqnarray}
&&[ (\Delta+ \eta(\gamma_s-\overline{\gamma}_s +2\mu_s +\overline{\mu}_s))(D-\eta(\rho_s+2\epsilon_s))-
(\overline{\delta} + \eta(\alpha_s +\overline{\beta}_s + 2 \pi_s - \overline{\tau}_s))(\delta -\eta(\tau_s + 2\beta_s))] \varphi_2 + \nonumber \\
&&- [\eta \nu_s\,D\,\varphi_1-\eta \lambda_s \delta\,\varphi_1 + 2\,\varphi_1\,\left((D +\eta(3 \epsilon_s+\overline{\epsilon}_s + \rho_s
- \overline{\rho}_s))\eta \nu_s
-2( \delta +\eta(\overline {\pi}_s + \tau_s -\overline{\alpha}_s+3\beta_s))\eta \lambda_s-4\eta \Psi_3\right)]  + \nonumber  \\
&&\varphi_0\,[(\Delta +\eta(\overline{\mu}_s-\overline{\gamma}_s+ \overline{\gamma}_s + 2\mu_s))\eta \lambda_s-(\overline{\delta}-\eta(\overline{\tau}_s
-\alpha_s 
- \overline{\beta}_s-2\pi_s))\eta \nu_s] +\eta \lambda_s\,\Delta\,\varphi_0 -\eta \nu_s\,\overline{\delta}\,\varphi_0=2\,\eta\,\hat \pi J_2. \label{eq:phi2a}
\end{eqnarray}
We perform a first order perturbation of the previous equations of the form 
$f\Rightarrow f+ 
f^{(1)}$. 
We restrict our analysis to background spacetimes where the 
quantities $\nu_s, \lambda_s, \kappa_s,\sigma_s, \varphi_0, \varphi_2$ are zero 
(as in the black hole family of spacetimes) to get
\begin{eqnarray}
&&[ (\Delta+\eta(\gamma_s-\overline{\gamma}_s +2\mu_s +\overline \mu_s))(D-\eta(\rho_s+2\epsilon_s))-
(\overline{\delta} + \eta(\alpha_s +\overline{\beta}_s + 2 \pi_s - \overline{\tau}_s))(\delta -\eta(\tau_s + 2\beta_s))]\varphi_2^{(1)} 
+ \label{eq:phi2b} \\
&&- \left[\eta {\nu_s}^{(1)}\,D\,\varphi_1-\eta {\lambda_s}^{(1)} \delta\,\varphi_1 + 2\,\varphi_1\,\left((D +\eta (3 \epsilon_s
+\overline{\epsilon}_s + \rho_s - \overline{\rho}))\eta {\nu_s}^{(1)}
- 2( \delta +\eta (\overline{\pi}_s + \tau_s -\overline{\alpha}_s+3\beta_s))\eta{\lambda_s}^{(1)} - \nonumber \right. \right. \\
&& \left. \left. 4\eta \Psi_3^{(1)}\right)\right] 
=2\,\eta\, \pi J_2^{(1)},\nonumber 
\end{eqnarray}
where we have kept only first order and background quantities. This 
expression can be simplified further by using Maxwell equations Eqs. 
(\ref{Muno}) $D\varphi_1=2\,\eta \rho_s\,\varphi_1 + 2\hat{\pi}\,J_l, \,$ 
and (\ref{Mdos}) $\delta \varphi_1=2\,\eta \tau_s\,\varphi_1 + 
2\hat{\pi}\,J_m$, to obtain
\begin{eqnarray}
&&[(\Delta + \eta(\overline{\mu}_s - \overline{\gamma}_s+\gamma_s + 2\mu_s))(D-\eta(\rho_s-2\epsilon_s)) - 
(\overline{\delta} - \eta(\overline{\tau}_s -\alpha_s - \overline{\beta}_s- 2 \pi_s))(\delta -\eta(\tau_s -2\beta_s))]\varphi_2^{(1)} = \nonumber \\
&&2 \eta \varphi_1[(D +\eta( 3 \epsilon_s+\overline{\epsilon}_s + 2\rho_s - \overline{\rho}_s)){\nu_s}^{(1)}
- (\delta+ \eta( \overline{\pi}_s  -\overline{\alpha}_s+3\beta_s+2\tau_s)){\lambda_s}^{(1)} - 
2 \Psi^{(1)}_3] + 2\,\eta\, \pi J_2^{(1)},
\label{eq:maxphi2}
\end{eqnarray}
with
\begin{eqnarray}
J_2^{(1)}&=&(\Delta+\eta(\gamma_s-\overline{\gamma}_s +2 \mu_s + \overline{\mu}_s))J_{\overline{m}}^{(1)}
- (\overline{\delta} +\eta(\alpha_s +\overline{\beta}_s + 2 \pi_s -\overline{\tau}_s)) J_n^{(1)}. 
\end{eqnarray}
Equation~(\ref{eq:maxphi2}) will be used later to eliminate the terms involving 
${\nu_s}^{(1)}$ and ${\lambda_s}^{(1)}$ in favor of the Weyl and Maxwell 
scalars.
It the next steps of the derivation it will be useful to commute the operators 
acting on $\varphi_2^{(1)}$, 
using
\begin{eqnarray}
&[\Delta, D ] =& \eta(\gamma_s +\overline{\gamma}_s)D+\eta(\epsilon_s + \overline{\epsilon}_s) \Delta - 
\eta(\overline{\tau}_s + \pi_s)\delta -\eta(\tau_s +\overline{\pi}_s)\overline{\delta}, \\
&[\overline{\delta}, \delta \small]  =& \eta(\overline{\mu}_s -\mu_s)D+\eta(\overline{\rho}_s-\rho_s)\Delta+
\eta(\alpha_s-\overline{\beta}_s)\delta+\eta({\beta}_s-\overline{\alpha}_s)\overline{\delta},
\end{eqnarray}
taking into account that 
\begin{eqnarray}
\eta D \mu_s&=& \eta \delta \pi_s +(\overline{\rho}_s \mu_s +\sigma_s \lambda_s)+\pi_s(\overline{\pi}-\overline{\alpha}_s+\beta_s)-
\mu_s(\epsilon_s+\overline{\epsilon}_s)-\nu_s \kappa_s +\eta \Psi_2 +2\Lambda, \\
\eta \overline{\delta}\,\beta_s&=& -\eta \delta \alpha_s-(\mu_s \rho_s-\lambda_s \sigma_s)-\alpha_s \overline{\alpha}_s -
\beta_s (\overline{\beta}_s +2\alpha_s) -\gamma_s(\rho_s-\overline{\rho}_s)-\epsilon_s(\mu_s -\overline{\mu}_s)
+\eta \Psi_2-\Phi_{11}-R,
\end{eqnarray} 
where $R$ is the Ricci scalar. After some algebra we arrive to the 
following expression relating 
$\varphi_2^{(1)}, \, {\nu_s}^{(1)}, \, {\lambda_s}^{(1)}, \, \Psi_3^{(1)},$ and the current $J_2^{(1)}$ (recalling that we are considering
a spacetime where the background quantities $\nu_s, \lambda_s, \kappa_s, \sigma_s, \varphi_0, \varphi_2$ are zero):
\begin{eqnarray}
&&[(D-\eta\,\left(\rho_s-3\epsilon_s-\overline{\epsilon}_s \right))(\Delta + \eta\,\left(2\mu_s + \overline{\mu}_s + 2 \gamma_s \right)) - 
(\delta + \eta\,\left(3\beta_s-\overline{\alpha}_s - \tau_s\right))(\overline{\delta} + \eta\,\left(2 \alpha_s
+ 2\pi_s - \tau_s\right))-3 \eta \Psi_2]\varphi_2^{(1)} 
= \nonumber \\
&&2 \eta \varphi_1[(D + \eta\,\left(2\rho_s - \overline{\rho}_s + 3 \epsilon_s+\overline{\epsilon}_s \right)){\nu_s}^{(1)}
- (\delta - \eta\,\left(\overline{\alpha}_s-3\beta_s-\overline{\pi}_s-2\tau_s\right)){\lambda_s}^{(1)} - 2 
\Psi^{(1)}_3] + 2\,\eta\, \pi J_2^{(1)}. 
\label{eq:maxphi2c}
\end{eqnarray}
It is important to notice that there is a typo in \cite{Chandrasekhar83} in 
the equivalent to this last
equation.

%%%%%%%%%%%%%%%%%%%%%%%%%%%%%%%%%%%%%%%%%%%%%%%%%%%%%%%%%%%%%%%%%%%%%%
\section{Perturbed Bianchi identities}
\label{sec:PBI}
%%%%%%%%%%%%%%%%%%%%%%%%%%%%%%%%%%%%%%%%%%%%%%%%%%%%%%%%%%%%%%%%%%%%%%
To derive the equations for the perturbations of $\Psi_4$ and $\Psi_3$, we 
start with the 
Bianchi identities: 
\begin{eqnarray}
&&(D+ \eta\,(4 \epsilon_s\, -\rho_s\,) )\Psi 
_{4}-(\overline{\delta}+2\eta\,(2\pi_s\, +\alpha_s\,) )\Psi _{3}
+\left(3\eta\, \Psi _{2}+2\Phi _{11}\right)\,\lambda_s \nonumber \\
&&=\eta\,(\overline{\delta}+2\eta\,(\alpha_s\, - \overline \tau_s))\Phi _{21}
-\eta\, \left( \Delta +\eta\,(\overline{\mu}_s + 2 \gamma_s\, - 2 \overline 
\gamma_s\, )\right)\Phi _{20}  \label{R321dB0_0}, \\ 
&&-(\delta +\eta\,(4\beta_s - \tau_s) )\Psi _{4}^{(1)}+(\Delta 
+ 2\eta\,( \gamma_s\, +2\mu_s\,) )\Psi _{3}-\left( 3 \eta\, \Psi_{2}-2\Phi 
_{11}\right) \nu_s  \nonumber \\
&&=\eta\, (\Delta+2\eta\,(\overline{\mu}_s+\gamma_s\,))\Phi _{21}
-\eta\,(\overline{\delta}+\eta\,(-\overline \tau_s+2\alpha_s\, 
+2\overline{\beta}_s))\Phi _{22}, \label{R321hB0_0} \\ \notag 
\end{eqnarray}
and with the Ricci identity:
\begin{eqnarray}
&&\Psi _{4}+  (\Delta +\eta\,(\mu_s\, +\overline{\mu}_s +3\gamma_s\, -
\overline{\gamma}_s))\,\lambda_s-(\overline{\delta}+\eta\,(3\alpha_s\, +\overline{\beta}_s+\pi_s\, -\overline{\tau}_s))\nu_s=0. \label{eq:BE40_0}
\end{eqnarray}
Performing a first order perturbation and considering spacetimes where the 
unperturbed quantities $\Psi_4, \Psi_3, \nu_s, \lambda_s, \sigma_s$ are zero, 
and the unperturbed sources $\Phi _{21}, \Phi _{20}, \Phi _{22}$ also vanish, 
one obtains:
\begin{eqnarray}
&&(D+ \eta\,(4 \epsilon_s\, -\rho_s\,) )\Psi _{4}^{(1)}-(\overline{\delta }+2\eta\,(2\pi_s\, +\alpha_s\,) )\Psi _{3}^{(1)}
+\left(3\eta\, \Psi _{2}+2\Phi _{11}\right)\,\lambda_s ^{(1)} \nonumber \\
&&=\eta\,(\overline{\delta}+2\eta\,(\alpha_s\, - \overline \tau_s))\Phi _{21}^{(1)}
-\eta\, \left( \Delta +\eta\,(\overline{\mu}_s + 2 \gamma_s\, - 2 \overline{\gamma}_s\, )\right)\Phi _{20}^{(1)},   \label{R321dB0} \\
&&-(\delta +\eta\,(4\beta_s - \tau_s) )\Psi _{4}^{(1)}+(\Delta 
+ 2\eta\,( \gamma_s\, +2\mu_s\,) )\Psi _{3}^{(1)}-\left( 3 \eta\, \Psi_{2}-2\Phi _{11}\right) \nu_s ^{(1)}  \nonumber \\
&&=\eta\, (\Delta+2\eta\,(\overline{\mu}_s+\gamma_s\,))\Phi _{21}^{(1)}
-\eta\,(\overline{\delta}+\eta\,(-\overline{\tau}_s+2\alpha_s\, ,
+2\overline{\beta}_s))\Phi _{22}^{(1)} \label{R321hB0}
\end{eqnarray}
and the perturbation of the Ricci identity \eqref{eq:BE40_0} gives,
\begin{equation}
\Psi _{4}^{(1)}+  (\Delta +\eta\,(\mu_s\, +\overline{\mu}_s+3\gamma_s\, -
\overline{\gamma}_s))\,\lambda_s^{(1)}-(\overline{\delta}+\eta\,(3\alpha_s\, +\overline{\beta}_s +\pi_s\, -\overline{\tau}_s))\nu_s^{(1)}=0. \label{eq:BE40}
\end{equation}
Eqs. (\ref{R321dB0})-(\ref{eq:BE40}) can be rewritten as
\begin{eqnarray}
{\cal O}_{1a}\,\Psi_{4}^{(1)} + {\cal O}_{1b}\,\Psi _{3}^{(1)} + \left(3\eta\, \Psi _{2} + 2\Phi _{11}\right)\,\lambda_s^{(1)} &=&
{\cal T}_{1a}\,\Phi _{21}^{(1)} + {\cal T}_{1b}\,\Phi _{20}^{(1)}, \label{R321dB} \\
{\cal O}_{2a}\,\Psi_{4}^{(1)} + {\cal O}_{2b}\,\Psi _{3}^{(1)} - \left(3\eta\, \Psi _{2} - 2\Phi _{11}\right)\,\nu_s^{(1)} &=&
{\cal T}_{2a}\,\Phi _{22}^{(1)} + {\cal T}_{2b}\,\Phi _{21}^{(1)}, \label{R321hB} \\
\Psi _{4}^{(1)} + {\cal O}_{4a}\,{\lambda_s}^{(1)} - {\cal O}_{4b}\,{\nu_s}^{(1)}&=&0, \label{eq:BE4} 
\end{eqnarray}

\noindent where we have defined
\begin{eqnarray}
{\cal O}_{3a}&=&(\Delta + \eta\,(3 \gamma_s -\overline \gamma_s + 4\mu_s 
+\overline \mu_s)), \qquad
{\cal O}_{3b}=(\overline \delta +\eta\,(- \overline  \tau_s + \overline 
\beta_s + 3 \alpha_s + 4 \pi_s)), \\ 
{\cal O}_{4a}&=&(\Delta +\eta\,(\mu_s\, +\overline{\mu}_s+3\gamma_s\, 
-\overline{\gamma}_s)), \qquad {\cal O}_{4b}=(\overline{\delta 
}+\eta\,(3\alpha_s\, +\overline{\beta}_s+\pi\, -\overline{\tau_s})),\nonumber \\
{\cal O}_{1a}&=&(D +\eta\,(4\epsilon_s\, - \rho_s\,) ), \qquad
\qquad \qquad \,\, {\cal O}_{2a}=-(\delta +\eta\,(4\beta_s - \tau_s) ),\nonumber \\
{\cal O}_{1b}&=&-(\overline{\delta }+2\eta\,(2\pi_s\, +\alpha_s\,) ),\qquad
\qquad \quad{\cal O}_{2b}=(\Delta + 2\eta\,( \gamma_s\, +2\mu_s\,) ), \nonumber \\
{\cal T}_{1a}&=&\eta\,(\overline{\delta}+2\eta\,(\alpha_s\, - \overline 
\tau_s)), \qquad
\qquad \qquad {\cal T}_{1b}=-\eta\, \left( \Delta +\eta\,(\overline{\mu}_s + 2 \gamma_s\, - 2 
{\overline \gamma}_s\, )\right),\nonumber \\
{\cal T}_{2a}&=&-\eta\, (\overline \delta +\eta\,(-\overline \tau_s+2\alpha\, 
+2\overline{\beta}_s)), \qquad
{\cal T}_{2b}=\eta\, (\Delta+2\eta\,(\overline{\mu}_s+\gamma_s\,)). 
\label{def:ops}
\end{eqnarray} 
%%%%%%%%%%%%%%%%%%%%%%%%%%%%%%%%%%%%%%%%%%55
The next commutation relation, obtained from a generalization of the Eq.~(\ref{eq:con_Dd}), see \cite{Newman62a, Chandrasekhar83}, 
will be useful to rewrite the 
Maxwell equations and the equation for the perturbations as second order 
differential equations,
\begin{equation}
[\Delta-q \mu_s+\overline{\mu}_s +(p+1) \gamma_s -\overline{\gamma}_s](\overline{\delta}+p\alpha_s-q\pi_s)
-[\overline{\delta}+\beta_s +(p+1)\alpha_s-q \pi_s- 
\overline{\tau}_s](\Delta-q\mu_s+p \gamma_s)=0, \label{identity2}
\end{equation}
where $p$ and $q$ are real numbers.
At this point it is possible to get an equation for $\Psi_4^{(1)}$ using the 
anti-commutation relation Eq.~(\ref{identity2}).
Applying the operator ${\cal O}_{3a}$ on Eq.~(\ref{R321dB}) 
and ${\cal O}_{3b}$ on Eq.~(\ref{R321hB}) and adding the resulting expressions we get:
\begin{eqnarray}
&&[{\cal O}_{3a}{\cal O}_{1a}+{\cal O}_{3b}{\cal O}_{2a}]\Psi_{4}^{(1)}
+[{\cal O}_{3a}{\cal O}_{1b}+{\cal O}_{3b}{\cal O}_{2b}]\Psi _{3}^{(1)}+{\cal O}_{3a}\left(3\eta\, \Psi _{2}+2\Phi _{11}\right){\lambda_s}^{(1)} 
- {\cal O}_{3b}\left( 3 \eta\, \Psi_{2}-2\Phi _{11}\right)\,{\nu_s}^{(1)}=\nonumber \\
&&[{\cal O}_{3a}{\cal T}_{1a}+{\cal O}_{3b}{\cal T}_{2b}]\Phi _{21}^{(1)}+[{\cal O}_{3a}{\cal T}_{1b}]\Phi _{20}^{(1)}
+[{\cal O}_{3b}{\cal T}_{2a}]\Phi _{22}^{(1)}, \label{eq:psi4}
\end{eqnarray}
furthermore, by using the anti-commutation relation  
$[{\cal O}_{3a}{\cal O}_{1b}+{\cal O}_{3b}{\cal O}_{2b}]=0$ we eliminate the term multiplying
$\Psi_3^{(1)}$.
We can further simplify this equation by means of the following relations:
${\cal O}_{3a}={\cal O}_{4a} + 3\,\eta\,\mu_s$, ${\cal O}_{3b}={\cal O}_{4b} + 3\,\eta\,\pi_s$, and 
\begin{eqnarray}
&&{\cal O}_{4a}(A\,{\lambda_s}^{(1)})=A\,{\cal O}_{4a}({\lambda_s}^{(1)}) + {\lambda_s}^{(1)}\,(\Delta\,A), \\  \nonumber
&&{\cal O}_{4b}(A\,{\nu_s}^{(1)})=A\,{\cal O}_{4b}({\nu_s}^{(1)}) + \nu^{(1)}\,(\overline{\delta}\,A), \nonumber
\end{eqnarray} 
for any function $A$. 
As a result, we can rewrite Eq.(\ref{eq:psi4}) as
\begin{eqnarray}
&&\left[{\cal O}_{3a}{\cal O}_{1a}+{\cal O}_{3b}{\cal O}_{2a}\right]\Psi _{4}^{(1)}
+ 3\eta\,\Psi _{2}\left[\left({\cal O}_{4a} + 3\,\eta\,\mu_s\right)\,{\lambda_s}^{(1)} - \left({\cal O}_{4b} + 3\,\eta\,\pi_s\right)\,
{\nu_s}^{(1)} \right] + 
{\lambda_s}^{(1)}\,\left[3\,\eta\,(\Delta\,\Psi _2) + 2\,(\Delta\,\Phi _{11})\right]   + \nonumber \\
&& 2\,\Phi _{11}\left[\left({\cal O}_{4a} + 3\,\eta\,\mu_s\right)\,{\lambda_s}^{(1)} + \left({\cal O}_{4b} + 3\,\eta\,\pi_s\right)\,{\nu_s}^{(1)} \right] - 
{\nu_s}^{(1)}\,\left[3\,\eta\,(\overline{\delta}\,\Psi_{2}) - 2\,(\overline{\delta}\,\Phi _{11})\right] =\nonumber \\
&&[{\cal O}_{3a}{\cal T}_{1a}+{\cal O}_{3b}{\cal T}_{2b}]\Phi _{21}^{(1)}+[{\cal O}_{3a}{\cal T}_{1b}]\Phi _{20}^{(1)}
+[{\cal O}_{3b}{\cal T}_{2a}]\Phi _{22}^{(1)}.
\end{eqnarray}
After collecting the terms that multiply at ${\lambda_s}^{(1)}$ and ${\nu_s}^{(1)}$ we get 
\begin{eqnarray}
&&[{\cal O}_{3a}{\cal O}_{1a}+{\cal O}_{3b}{\cal O}_{2a}]\,\Psi _{4}^{(1)}
+ 3\eta\,\Psi _{2}\left[{\cal O}_{4a}\,{\lambda_s}^{(1)} - {\cal O}_{4b}\,{\nu_s}^{(1)} \right] + 
{\lambda_s}^{(1)}\,\left[3\,\eta\,(\Delta\,\Psi _2) + 9\,\mu_s\,\Psi_2 + 2\,(\Delta\,\Phi _{11}) + 6\,\eta\,\mu_s\,\Phi_{11}\right]   + \nonumber \\
&& 2\,\Phi _{11}\left[{\cal O}_{4a}\,{\lambda_s}^{(1)} + {\cal O}_{4b}\,{\nu_s}^{(1)} \right] - 
{\nu_s}^{(1)}\,\left[3\,\eta\,(\overline{\delta}\,\Psi_{2}) + 9\,\pi_s\,\Psi_2 - 2\,(\overline{\delta}\,\Phi _{11}) -
6\,\eta\,\pi_s\,\Phi_{11}\right] =\nonumber \\
&&[{\cal O}_{3a}{\cal T}_{1a}+{\cal O}_{3b}{\cal T}_{2b}]\Phi _{21}^{(1)}+[{\cal O}_{3a}{\cal T}_{1b}]\Phi _{20}^{(1)}
+[{\cal O}_{3b}{\cal T}_{2a}]\Phi _{22}^{(1)}.
\end{eqnarray}
Using Eq.~(\ref{eq:BE4}), ${\cal O}_{4a}\,{\lambda_s}^{(1)} - {\cal O}_{4b}\,{\nu_s}^{(1)}=-\Psi _{4}^{(1)}$, in the 
second and fourth terms, we obtain:
\begin{eqnarray}
&&[{\cal O}_{3a}{\cal O}_{1a}+{\cal O}_{3b}{\cal O}_{2a} - 3\eta\,\Psi _{2} + 2\,\Phi _{11}]\,\Psi _{4}^{(1)} +
{\lambda_s}^{(1)}\,\left[3\,\eta\,(\Delta\,\Psi _2) + 9\,\mu_s\,\Psi_2 + 2\,(\Delta\,\Phi _{11}) + 6\,\eta\,\mu_s\,\Phi_{11}\right]   + 
4\,\Phi _{11}\,{\cal O}_{4a}\,{\lambda_s}^{(1)} + \nonumber \\
&&  
{\nu_s}^{(1)}\,\left[3\,\eta\,(\overline{\delta}\,\Psi_{2}) + 9\,\pi_s\,\Psi_2 - 2\,(\overline{\delta}\,\Phi _{11}) -
6\,\eta\,\pi_s\,\Phi_{11}\right] =[{\cal O}_{3a}{\cal T}_{1a}+{\cal O}_{3b}{\cal T}_{2b}]\Phi _{21}^{(1)}+[{\cal O}_{3a}{\cal T}_{1b}]\Phi _{20}^{(1)}
+[{\cal O}_{3b}{\cal T}_{2a}]\Phi _{22}^{(1)}. \label{eq:lambdanupert}
\end{eqnarray}
Using the unperturbed Einstein and Maxwell equations:
\begin{eqnarray}
\Delta\,\Psi _{2}&=& - 3\eta\,\mu_s\,\Psi_{2} - 2\,\mu_s\,\Phi_{11}, \hspace{1cm} \Delta\,\Phi _{11}=- 4\,\mu_s\,\eta\,\Phi_{11}, \label{eq:Delta_A} \\
\overline{\delta}\,\Psi _{2}&=&- 3\eta\,\pi_s\,\Psi_{2} + 2\,\pi_s\,\Phi_{11}, \hspace{1cm}  
\overline{\delta}\,\Phi _{11}= -4\,\eta\,\pi_s\,\Phi_{11},\label{eq:deltac_A}
\end{eqnarray}
Eq.~(\ref{eq:lambdanupert}) becomes:
\begin{eqnarray}
&&[{\cal O}_{3a}{\cal O}_{1a}+{\cal O}_{3b}{\cal O}_{2a} - 3\eta\,\Psi _{2} + 2\,\Phi _{11}]\,\Psi _{4}^{(1)} 
+ 4\,\Phi _{11}\,\left({\cal O}_{4a} - 2\,\eta\,\mu_s\right)\,{\lambda_s}^{(1)} - 8\,\eta\,\pi_s\,\Phi_{11}\,{\nu_s}^{(1)} =\nonumber \\
&& 
[{\cal O}_{3a}{\cal T}_{1a}+{\cal O}_{3b}{\cal T}_{2b}]\Phi _{21}^{(1)}+[{\cal O}_{3a}{\cal T}_{1b}]\Phi _{20}^{(1)}
+[{\cal O}_{3b}{\cal T}_{2a}]\Phi _{22}^{(1)}. \label{eq:pert_gral}
\end{eqnarray}

It is straightforward to see that in vacuum, the usual perturbation 
equation for $\Psi_4^{(1)}$ is recovered \cite{Degollado:2009rw}.  
In the present case, by means of the 
Eqs.~(\ref{R321dB})-(\ref{R321hB}) it can be seen that we have three
unknowns: $\Psi_4^{(1)}$, $\Psi_3^{(1)}$, and the perturbed electromagnetic 
function, $\varphi_2^{(1)}$. In order to solve for these perturbations it is 
necessary to find another 
equation for ${\Psi_3}^{(1)}$. This can be done by 
performing a 
similar procedure as explained above, but now acting on 
Eqs.~(\ref{R321dB})-(\ref{R321hB}) to eliminate the operator acting on 
${\Psi_4}^{(1)}$.
The operators that are convenient to achieve that goal are:
\begin{equation}
{\cal O}_{5a}=(D + \eta (3 \epsilon_s +\overline \epsilon_s-\rho_s - \overline \rho_s)),  
\hspace{1cm} {\cal O}_{5b}=(\delta +\eta (3  \beta_s - \overline  \alpha_s - \tau_s + \overline \pi_s)).
\end{equation}

We also need to consider the action of the operators $D$ and 
$\delta$ on $\Psi_2$ and on $\Phi_{11}$:
\begin{equation}
D \Psi _{2}=  3\eta \rho_s  \Psi_{2} + 2  \rho_s \Phi_{11},   \hspace{1cm}   D\Phi_{11}=4\,\eta \rho_s \Phi_{11},  \hspace{1cm} \, 
\delta \Psi _{2}= 3\eta \tau_s  \Psi_{2} - 2 \tau_s \Phi_{11}, \hspace{1cm}   \delta \Phi_{11}=2 \eta \tau_s \Phi_{11}.
\end{equation}
Applying ${\cal O}_{5b}$ on Eq.~(\ref{R321dB}), and
${\cal O}_{5a}$ on Eq.~(\ref{R321hB}), and adding the resulting equations, by 
means of the commuting operator (\ref{identity2}),
we obtain:
\begin{eqnarray}
&&[{\cal O}_{5b}{\cal O}_{1b}+{\cal O}_{5a}{\cal O}_{2b}]\Psi _{3}^{(1)}+\left(3\eta \Psi_2+2\,\Phi_{11}\right)\,{\cal O}_{5b}\,\lambda^{(1)} - 
{\cal O}_{5a}\,\left(3\eta \Psi_2-2\,\Phi_{11}\right)\,\nu^{(1)} - [{\cal O}_{5b}{\cal T}_{1a}+{\cal O}_{5a}{\cal T}_{2b}]\,\Phi _{21}^{(1)} \nonumber \\
&& = [{\cal O}_{5b}{\cal T}_{1b}]\Phi _{20}^{(1)} + [{\cal O}_{5a}{\cal T}_{2a}]\Phi _{22}^{(1)},
\end{eqnarray}
which implies
\begin{eqnarray}
&&[{\cal O}_{5b}{\cal O}_{1b}+{\cal O}_{5a}{\cal O}_{2b}]\Psi _{3}^{(1)} + 3\eta \Psi_2\,[{\cal O}_{5b}\,{\lambda_s}^{(1)} 
- \left({\cal O}_{5a} + \eta\,\rho_s\right)\,{\nu_s}^{(1)}]  + 2\,\Phi_{11}\,[{\cal O}_{5b}\,{\lambda_s}^{(1)} 
+ \left({\cal O}_{5a} + \eta\,\rho_s\right)\,{\nu_s}^{(1)}]  -  \nonumber \\
&& 6\,\rho_s\,\Psi_2\,{\nu_s}^{(1)} - [{\cal O}_{5b}{\cal T}_{1a}+{\cal O}_{5a}{\cal T}_{2b}]\,\Phi _{21}^{(1)} =
[{\cal O}_{5b}{\cal T}_{1b}]\,{T_{20}}^{(1)} + [{\cal O}_{5a}{\cal T}_{2a}]\,{T_{22}}^{(1)}. \label{eq:gral1}
\end{eqnarray}

In order to get rid of the perturbed spinor coefficients, we will make 
use of the Maxwell relation, Eq.~(\ref{eq:maxphi2c}).
Additionally, we will make two 
simplifying assumptions. The first one 
is that the spacetime is Reissner-N\"ordstrom, and the 
second, following Chandrasekhar \cite{Chandrasekhar83}, we will use the so 
called {\it phantom gauge}, where, using the fact
that $\chi=2\,\varphi_1\,{\Psi_3}^{(1)} - 3\,\Psi_2\,{\varphi_2}^{(1)}$ is 
invariant under gauge transformations, one 
can choose a gauge where ${\varphi_2}^{(1)}=0$. Using this gauge we will be 
able to find an equation for $\Psi_3^{(1)}$.
In order to take into account the contribution of the different components of 
the matter content we will consider that the perturbation for the matter fields 
can be expressed 
as 
\begin{equation}
\Phi_{21}^{(1)}={\Phi_{21}^{(1)}}_{\rm elec} + T_{21}^{(1)}, \hspace{1cm} \Phi_{20}^{(1)}={\Phi_{20}^{(1)}}_{\rm elec} + T_{20}^{(1)},
\hspace{1cm} \Phi_{22}^{(1)}={\Phi_{22}^{(1)}}_{\rm elec} + T_{22}^{(1)}, \label{eq:_split}
\end{equation}
where $T_{ab}^{(1)}$ is the term for external matter that perturbs the 
background. 
Since in the Reissner-N\"ordstrom background $\varphi_2=0$ and 
$\varphi_0 = 0$, we have that the three perturbed components of the Ricci tensor 
due to
the perturbed electromagnetic field, are zero:
\begin{equation}
 {\Phi_{21}^{(1)}}_{\rm elec}=\varphi_2^{(1)}\varphi_1 + \varphi_2\varphi_1^{(1)}  = \varphi_2^{(1)}\varphi_1, \qquad {\rm and}\qquad
 \Phi_{20}^{(1)}= \varphi_2^{(1)}\varphi_0 + \varphi_2\varphi_0^{(1)} = 0,
\end{equation}
and ${\Phi_{22}^{(1)}}_{\rm elec}=0$.

In the next section we will show that, in the Reissner-N\"ordstrom spacetime, the 
spin coefficients are real and $\pi_s= 0= \tau_s= \gamma_s$ and
$\alpha_s=-\beta_s$.
Furthermore, the following relations are valid
${\cal O}_{4a} - 2\,\eta\,\mu_s=\Delta$, and, ${\cal O}_{3a}{\cal T}_{1a}+{\cal O}_{3b}{\cal T}_{2b}=2\,\eta\,\left(\Delta + 
4\,\eta\,\mu_s\right)\,\left(\overline{\delta} + 2\,\eta\,\alpha_s\right)$. 
Finally, in the forthcoming analysis we will use the signature $(-,+,+,+)$
so that $\eta=-1$.

Using the phantom gauge, the Maxwell relation 
Eq.~(\ref{eq:maxphi2c}) takes the form
\begin{equation}
{\cal O}_{5b}\,{\lambda_s}^{(1)} - \left({\cal O}_{5a} - 3\,\rho\right)\,{\nu_s}^{(1)}
= - 2\,{\Psi_3}^{(1)} + \frac{\pi}{ \varphi_1} J^{(1)}_2.
\label{eq:d_Maxwell_pha1}
\end{equation}

The first coupled equations for ${\Psi_4}^{(1)}$ and ${\lambda_s}^{(1)}$, Eq.~(\ref{eq:pert_gral}), takes the form:
\begin{eqnarray}
&&
[{\cal O}_{3a}{\cal O}_{1a}+{\cal O}_{3b}{\cal O}_{2a} + 3\,\Psi _{2} + 2\,\Phi _{11}]\,\Psi _{4}^{(1)} 
+ 4\,\Phi _{11}\,\left({\cal O}_{4a} + 2\,\mu_s\right)\,{\lambda_s}^{(1)} =
\nonumber \\ && 
[{\cal O}_{3a}{\cal T}_{1a}+{\cal O}_{3b}{\cal T}_{2b}]\,T_{21}^{(1)}+[{\cal O}_{3a}{\cal T}_{1b}]\,T _{20}^{(1)}
+[{\cal O}_{3b}{\cal T}_{2a}]\,T_{22}^{(1)},  \label{eq:pert_4-3_RN}
\end{eqnarray}
that is
%
%\begin{eqnarray}
\begin{equation}
%&&
[{\cal O}_{3a}{\cal O}_{1a}+{\cal O}_{3b}{\cal O}_{2a} + 3\Psi _{2} + 2\,\Phi _{11}]\,\Psi _{4}^{(1)} 
+ 4\,\Phi _{11}\,\Delta\,\lambda^{(1)}  =
%\nonumber \\ && 
-2\,\left(\Delta - 4\,\mu_s\right)\,\left(\overline{\delta} + 2\,\beta_s\right)\,T_{21}^{(1)}
+[{\cal O}_{3a}{\cal T}_{1b}]\,T_{20}^{(1)} + [{\cal O}_{3b}{\cal T}_{2a}]\,T_{22}^{(1)}. \label{eq:psi4_lambda}
\end{equation}
%\end{eqnarray}
%

Substituting Eq.~(\ref{eq:d_Maxwell_pha1}) in Eq.~(\ref{eq:gral1}), we 
obtain a second coupled equation for 
${\Psi_3}^{(1)}$ and the perturbed spinor coefficients ${\nu_s}^{(1)}$, and ${\lambda_s}^{(1)}$:
\begin{eqnarray}
&&\left[{\cal O}_{5b}\,{\cal O}_{1b} + {\cal O}_{5a}\,{\cal O}_{2b} + 2\,\left(3\,\Psi_2 + 2\,\Phi_{11}\right)\right]\Psi _{3}^{(1)} + 
4\,\Phi_{11}\,{\cal O}_{5b}\,{\lambda_s}^{(1)}  + 4\,\rho_s\,\Phi_{11}\,{\nu_s}^{(1)} =  \nonumber \\
&&
[{\cal O}_{5b}{\cal T}_{1a}+{\cal O}_{5a}{\cal T}_{2b}]\,{T_{12}}^{(1)} +  
[{\cal O}_{5b}{\cal T}_{1b}]\,{T_{20}}^{(1)} + [{\cal O}_{5a}{\cal T}_{2a}]\,{T_{22}}^{(1)}
+ \frac{ \pi}{\varphi_1}\left(3\,\Psi_2 + 2\,\Phi_{11}\right)J_{2}^{(1)}. 
\label{eq:psi3_nu_lambda}
\end{eqnarray}
The task now is to get rid of the perturbed spinor coefficients. This 
is done from  
Eq.~(\ref{R321dB}), solving for ${\lambda_s}^{(1)}$ and from Eq.~(\ref{R321hB})
solving for ${\nu_s}^{(1)}$
\begin{eqnarray}
{\lambda_s}^{(1)}&=& \frac{1}{2\Phi_{11} - 3\,\Psi_2}\left[ -{\cal O}_{1a} \Psi _{4}^{(1)}-{\cal O}_{1b} \Psi _{3}^{(1)}
+{\cal T}_{1a}\,T_{21}^{(1)}+{\cal T}_{1b}\,T_{20}^{(1)}\right],
\label{eq:lambda_p} \\
{\nu_s}^{(1)} &=& -\frac{1}{(3\,\Psi_2+2\Phi_{11})}\left[ {\cal O}_{2a} \Psi _{4}^{(1)}
+{\cal O}_{2b} \Psi _{3}^{(1)}-{\cal T}_{2a}\,T_{22}^{(1)}-{\cal T}_{2b}\,T_{21}^{(1)}\right]. \label{eq:nu_p}
\end{eqnarray}

Thus, we can express the term $4\,\Phi _{11}\,\Delta\,{\lambda_s}^{(1)}$ 
in~(\ref{eq:psi3_nu_lambda}) as: 
\begin{equation}
4\,\Phi _{11}\,\Delta\,{\lambda_s}^{(1)}= \frac{4\,\Phi _{11}}{2\Phi_{11} - 3\,\Psi_2}\,
\left(\Delta - \frac{14\,\Phi_{11} - 9\,\Psi_2}{2\Phi_{11} - 3\,\Psi_2}\,\mu_s\right)\,
\left[- {\cal O}_{1a} \Psi _{4}^{(1)}-{\cal O}_{1b} \Psi _{3}^{(1)}
+{\cal T}_{1a}\,T_{21}^{(1)}+{\cal T}_{1b}\,T_{20}^{(1)}\right], \label{eq:phi11_lambda}
\end{equation}
where we have used Eq.~(\ref{eq:Delta_A}).

After substituting expression~(\ref{eq:phi11_lambda}) on Eq.~(\ref{eq:psi4_lambda}) we get:
\begin{eqnarray}
&&\left\{\left[{\cal O}_{3a} - \frac{4\,\Phi _{11}}{2\Phi_{11} - 3\,\Psi_2}\,
\left(\Delta - \frac{14\,\Phi_{11} - 9\,\Psi_2}{2\Phi_{11} - 3\,\Psi_2}\,\mu_s\right) \right]{\cal O}_{1a}
+ {\cal O}_{3b}{\cal O}_{2a} + 3\,\Psi _{2} + 2\,\Phi _{11}\right\}\,\Psi _{4}^{(1)} - \nonumber \\
&& \frac{4\,\Phi _{11}}{2\Phi_{11} - 3\,\Psi_2}\,
\left(\Delta - \frac{14\,\Phi_{11} - 9\,\Psi_2}{2\Phi_{11} - 3\,\Psi_2}\,\mu_s\right)\,
{\cal O}_{1b} \Psi _{3}^{(1)}= \nonumber \\
&& -\left[\frac{4\,\Phi _{11}}{2\Phi_{11} - 3\,\Psi_2}\,
\left(\Delta - \frac{14\,\Phi_{11} - 9\,\Psi_2}{2\Phi_{11} - 3\,\Psi_2}\,\mu_s\right)\,{\cal T}_{1a}
+ 2\,\left(\Delta - 4\,\eta\,\mu_s\right)\,\left(\overline{\delta} + 2\,\beta_s\right)\right]\,T_{21}^{(1)}
 \nonumber \\
&& + \left[{\cal O}_{3a} - \frac{4\,\Phi _{11}}{2\Phi_{11} - 3\,\Psi_2}\,
\left(\Delta - \frac{14\,\Phi_{11} - 9\,\Psi_2}{2\Phi_{11} - 3\,\Psi_2}\right)\,\mu_s\right]\,{\cal T}_{1b}\,T_{20}^{(1)}
 + [{\cal O}_{3b}{\cal T}_{2a}]\,T_{22}^{(1)}.
\end{eqnarray}
Writing explicitly the radial-temporal operators and after some algebraic steps we obtain 
\begin{eqnarray}
&&\left\{\left[\Delta + 
\chi
\,\mu_s\,
\left(5 - \frac{4\,\Phi_{11}\,\left(14\,\Phi_{11} - 
9\,\Psi_2\right)}{\left(2\Phi_{11} - 
3\,\Psi_2\right)^2}\right) \right]\,(D + \rho_s - 4\epsilon_s) 
- 
%\chi\frac{2\Phi_{11} - 3\,\Psi_2}{2\Phi_{11} + 3\,\Psi_2}
\chi
\,
{\cal O}_{3b}{\cal O}_{2a}  - 2\,\Phi _{11} + 3\,\Psi _{2}
\right\}\,\Psi _{4}^{(1)} \nonumber \\
&& -\frac{4\,\Phi _{11}}{2\Phi_{11} + 3\,\Psi_2}\,
\left(\Delta - \frac{14\,\Phi_{11} - 9\,\Psi_2}{2\Phi_{11} - 
3\,\Psi_2}\,\mu_s\right)\,{\cal O}_{1b}\,\Psi _{3}^{(1)}
=\nonumber \\
&& -\left[\frac{6\,\Psi _{2}}{2\Phi_{11} + 3\,\Psi_2}\,\Delta + 
4\,
\chi
\,\left(2 - 
\frac{\Phi_{11}\,\mu_s\,\left(14\,\Phi_{11} - 9\,
\Psi_2\right)}{\left(2\Phi_{11} - 3\,\Psi_2\right)^2}\right)\right]\,
\left(\overline{\delta} + 2\,\beta_s\right)\,T_{21}^{(1)}  \nonumber \\
&& + \left[\Delta + 
%\frac{2\Phi_{11} - 3\,\Psi_2}{2\Phi_{11} + 3\,\Psi_2}
\chi
\,
\mu_s\left(5 - \frac{4\,\Phi_{11}\,\left(14\,\Phi_{11} - 
9\,\Psi_2\right)}{\left(2\Phi_{11} - 3\,\Psi_2\right)^2}\right) \right]\,
\left( \Delta - \mu_s\right)\,T _{20}^{(1)}
 - 
% \frac{2\Phi_{11} - 3\,\Psi_2}{2\Phi_{11} + 3\,\Psi_2}
 \chi
 \,
[{\cal O}_{3b}{\cal T}_{2a}]\,T_{22}^{(1)}, \label{eq:psi4_matter}
\end{eqnarray}
which is the first coupled equation for $\Psi _{4}^{(1)}$ and $\Psi _{3}^{(1)}$, with sources $T_{21}^{(1)}, T_{20}^{(1)}$
and $T_{22}^{(1)}$ and we have defined:
\begin{equation}\label{eq_chidef}
 \chi:=\frac{2\Phi_{11} - 3\,\Psi_2}{2\Phi_{11} + 3\,\Psi_2}.
\end{equation}

Regarding the second equation needed to close the system
we substitute ${\lambda_s}^{(1)}$ given by Eq.~(\ref{eq:lambda_p}) and 
${\nu_s}^{(1)}$ given by Eq.~(\ref{eq:nu_p}), 
in Eq.~(\ref{eq:psi3_nu_lambda}) to get:
\begin{eqnarray}
&&\left[\left(1 + 4\,\frac{\Phi_{11}}{3\,\Psi_2 - 2\,\Phi_{11}}\right)\,{\cal O}_{5b}{\cal O}_{1b} + 
\left({\cal O}_{5a} - 4\,\rho_s\,\frac{\Phi_{11}}{3\,\Psi_2 + 2\,\Phi_{11}}\right)\,{\cal O}_{2b} + 
2\,\left(3\,\Psi_2 + 2\,\Phi_{11}\right)\right]\,\Psi _{3}^{(1)}  \nonumber \\
&&
+ \frac{4\,\Phi_{11}}{3\,\Psi_2 - 2\,\Phi_{11}}\,
\left[{\cal O}_{5b}{\cal O}_{1a} + 
\rho_s\,\chi
\,{\cal O}_{2a}\right]\,\Psi _{4}^{(1)} = \nonumber \\
&&+ \left[\left(1 + \frac{4\,\Phi_{11}}{3\,\Psi_2 - 2\,\Phi_{11}}\right)\,{\cal O}_{5b}{\cal T}_{1a} + \left({\cal O}_{5a} 
- 4\,\rho_s\,\frac{\Phi_{11}}{3\,\Psi_2 + 2\,\Phi_{11}}\right)\,{\cal T}_{2b}\right]\,{T_{12}}^{(1)}  \nonumber \\
&& 
+ \left(1 + \frac{4\,\Phi_{11}}{3\,\Psi_2 - 2\,\Phi_{11}}\right)\,{\cal O}_{5b}{\cal T}_{1b}\,{T_{20}}^{(1)} +
\left({\cal O}_{5a} - 4\,\rho_s\,\frac{\Phi_{11}}{3\,\Psi_2 + 2\,\Phi_{11}}\right)\,{\cal T}_{2a}\,{T_{22}}^{(1)}
+ \frac{\pi}{\varphi_1}\left(3\,\Psi_2 + 2\,\Phi_{11}\right)J_{2}^{(1)} . 
\label{eq:general1}
\end{eqnarray}
In order to proceed further, we will specify the derivation for the Reissner-N\"ordstrom spacetime in a particular coordinate system.

%%%%%%%%%%%%%%%%%%%%%%%%%%%%%%%%%%%%%%%%%%%%%
%\section{Reissner N\"ordstrom metric}
%\label{sec:RN}
%%%%%%%%%%%%%%%%%%%%%%%%%%%%%%%%%%%%%%%%%%%%%

%%%%%%%%%%%%%%%%%%%%%%%%%%%%%%%%%%%%%%%%%%%%%%%%%%%%%%%%%%%%%%%%%%%%%%%%%%
\section{Coupled equations with sources in Reissner-N\"ordstrom, ghost gauge}
\label{sec:CES_RN}
%%%%%%%%%%%%%%%%%%%%%%%%%%%%%%%%%%%%%%%%%%%%%%%%%%%%%%%%%%%%%%%%%%%%%%%%%%%%%

The element of line for the Reissner N\"ordstrom spacetime in Kerr Schild-type 
coordinates is 
\begin{equation}
ds^{2}=-\left(1-\frac{2\,M}{r}+\frac{Q^2}{r^2}\right)dt^{2}+2\left(\frac{2\,M}{r}-
\frac{Q^2}{r^2}\right)dtdr+\left(1+\frac{2\,M}{r}-\frac{Q^2}{r^2}\right)dr^{2}+
r^{2}\left( d\theta ^{2}+\sin ^{2}\theta d\phi^{2}\right).  \label{eq:metric}
\end{equation}

In these coordinates we choose a null tetrad:
\begin{align}
l^{\mu}& =\frac{1}{2}\left( 1+\frac{2\,M}{r}-\frac{Q^2}{r^2}, 1-\frac{2\,M}{r}+\frac{Q^2}{r^2}, 0, 0\right),  \notag \\
k^{\mu}& = \left( 1,  -1, 0, 0\right),  \notag \\
m^{\mu}& =\frac{1}{\sqrt{2}\,r}\left( 0,0, 1, i\,\csc{\theta}  \right). 
%\overline{m}^{\mu}& =\frac{1}{\sqrt{2}\,r}\left( 0,0, 1, -i\,\csc{\theta}  \right),  
\end{align}
The only non-vanishing components of the Weyl and Ricci scalars are
\begin{equation}
\Psi _{2}=\frac{M}{r^3} -\frac{Q^2}{r^4}, \quad \Phi _{11}=\frac{Q^2}{2\,r^4},
\label{weyl}
\end{equation}
whereas the non-zero spin coefficients are 
\begin{equation}
\mu_s  =\frac{1}{r}, \quad 
\rho_s =\frac{r^2-2Mr+Q^2}{2 r^3}, \quad
\epsilon_s = -\frac{1}{2}\left(\frac{M}{r^2}-\frac{Q^2}{r^3}\right), \quad 
\beta_s  =-\alpha_s =-\frac{1}{2\sqrt{2}}\frac{\cot \theta }{r}. \quad
\label{spinc}
\end{equation}
%%%%%%%%%%%%%%%%%%%%%%%%%%%%%%%%%%%%%%%%%%%%%%%%%%%%%%%%%%%%%%%%%%%%%%%%%%%%%%
In this coordinate system we will write explicit equations for the 
perturbations $\Psi^{(1)}_{4}$ and $\Psi^{(1)}_{3}$. From 
Eq.~(\ref{eq:general1}), 
simplifying, and using that ${\cal O}_{5b}=-{\cal O}_{2a}$ in Reissner-N\"ordstrom, and that ${\cal O}_{5b}\,{\cal O}_{1a}=
\left({\cal O}_{1a}+\,\rho_s\right)\,{\cal O}_{5b}=-\left({\cal 
O}_{1a}+\,\rho_s\right)\,{\cal O}_{2a}$,
we obtain the second equation for $\Psi _{3}^{(1)}$, and $\Psi _{4}^{(1)}$ with sources:
\begin{eqnarray}
&&\left[\left({\cal O}_{5a} - 4\,\rho_s\,
\frac{\Phi_{11}}{3\,\Psi_2 + 2\,\Phi_{11}}\right)\,{\cal O}_{2b} +
%\frac{3\,\Psi_2 + 2\,\Phi_{11}}{3\,\Psi_2 - 2\,
%\Phi_{11}}
\frac{1}{\chi}
\,{\cal O}_{5b}{\cal O}_{1b} +  
2\,\left(3\,\Psi_2 + 2\,\Phi_{11}\right)\right]\,\Psi _{3}^{(1)}  \nonumber \\
&&
- \frac{4\,\Phi_{11}}{3\,\Psi_2 - 2\,\Phi_{11}}\,\left({\cal O}_{1a} + 
6\,\rho_s\,\frac{\Psi_2}{3\,\Psi_2 + 2\,\Phi_{11}}\right)\,{\cal O}_{2a}\,
\Psi _{4}^{(1)} = 
\left[\left({\cal O}_{5a} - 4\,\rho_s\,\frac{\Phi_{11}}{3\,\Psi_2 + 
2\,\Phi_{11}}\right)\,{\cal T}_{2b} +
%\frac{3\,\Psi_2 + 2\,\Phi_{11}}{3\,\Psi_2 - 2\,\Phi_{11}}
\frac{1}{\chi}
\,
{\cal O}_{5b}{\cal T}_{1a}\right]\,{T_{12}}^{(1)}  \nonumber \\
&& 
%+ \frac{3\,\Psi_2 + 2\,\Phi_{11}}{3\,\Psi_2 - 2\,\Phi_{11}}
\frac{1}{\chi}
\,{\cal O}_{5b}{\cal T}_{1b}\,{T_{20}}^{(1)} +
\left({\cal O}_{5a} - 4\,\rho_s\,\frac{\Phi_{11}}{3\,\Psi_2 + 2\,
\Phi_{11}}\right)\,{\cal T}_{2a}\,{T_{22}}^{(1)}
+ \frac{ \pi}{\varphi_1}\left(3\,\Psi_2 + 2\,\Phi_{11}\right)J_{2}^{(1)}. 
\label{eq:psi3_matter}
\end{eqnarray}
Regarding the source of the perturbation, 
let us assume that such matter source is a
cloud of charged particles with charge $e$ and mass $m$ that behave like dust 
(pressure less fluid).
The stress energy tensor is
\begin{equation}
T_{\mu\nu} = \rho u_{\mu}u_{\nu},
\end{equation}
$\rho$ is the rest mass density and $u^{\mu}$ is the four velocity. 
Furthermore, in our analysis we will consider 
that the fluid is falling radially into the black hole with four velocity:
\begin{equation}
u^{\mu} = [u^{t}(t,r),u^{r}(t,r),0,0]\ .
\end{equation}
Let us further assume that the particles have a constant
charge-mass ratio $q=(e/m)$ throughout the cloud. Then, the electric current induced by the motion of the particles is
\begin{equation}
J^{\mu}_{{\rm el}} = q\rho u^{\mu}.
\end{equation}
We will assume that this current is the source for the electromagnetic field in 
Maxwell equations namely, ${J^{\mu}}^{(1)}=J^{\mu}_{{\rm el}}$.
With the previous assumptions, the non-vanishing projections of the stress 
energy tensor and electric current are
$T _{22}^{(1)}$ and ${J_2}^{(1)}$. The perturbation equations, 
Eq.~(\ref{eq:psi4_matter}) and Eq.~(\ref{eq:psi3_matter}) become
%
%\small{

\begin{eqnarray}
&&\left\{\left[\Delta + \mu_s\, \chi
%\frac{3\,\Psi_2 - 2\Phi_{11}}{2\Phi_{11} + 
%3\,\Psi_2}\,
\left(5 + \frac{4\,\Phi_{11}\,\left(9\,\Psi_2 - 14\,
\Phi_{11}\right)}{\left(3\,\Psi_2 - 2\Phi_{11} 
\right)^2}\right) \right]\,(D +\rho_s - 4\epsilon_s) 
+ 
\chi
%\frac{3\,\Psi_2 - 2\Phi_{11}}{3\,\Psi_2 + 2\Phi_{11}}
\,
{\cal O}_{3b}{\cal O}_{2a}  - 2\,\Phi _{11} + 3\,\Psi _{2}
\right\}\,\Psi _{4}^{(1)}  \nonumber \\
&& -\frac{4\,\Phi _{11}}{2\Phi_{11} + 3\,\Psi_2}\,
\left(\Delta - \frac{9\,\Psi_2 - 14\,\Phi_{11}}{3\,\Psi_2 - 2\Phi_{11}}\,
\mu_s\right)\,{\cal O}_{1b}\,\Psi _{3}^{(1)}
= -4\, \pi\,
%\frac{3\,\Psi_2 - 2\Phi_{11}}{3\,\Psi_2 + 2\Phi_{11}}
\chi 
\,[{\cal 
O}_{3b}{\cal T}_{2a}]\,T_{22}^{(1)}.
\label{eq:psi4_radang}
\end{eqnarray}
%}

%\small{
\begin{eqnarray}
&&\left[\left(D +2\,\rho_s - 4\,\epsilon_s - 4\,\rho_s\,
\frac{\Phi_{11}}{3\,\Psi_2 + 2\,\Phi_{11}}\right)\,{\cal O}_{2b}  +
%\left(\frac{3\,\Psi_2 + 2\,\Phi_{11}}{3\,\Psi_2 - 2\,\Phi_{11}}\right)\,
\frac{1}{\chi}
{\cal O}_{5b}{\cal O}_{1b} 
+ 6\,\Psi_2 + 4\,\Phi_{11} \right]\,\Psi _{3}^{(1)}  \nonumber \\ 
&& - 4\,\frac{\Phi_{11}}{3\,\Psi_2 - 2\,\Phi_{11}}
\left[D + \rho_s - 4\,\epsilon_s + 
6\,\rho_s\,\frac{\Psi_2}{3\,\Psi_2 + 2\,\Phi_{11}}\right]\,{\cal O}_{2a}\,
\Psi _{4}^{(1)}  = 
\nonumber \\
&& 
4\, \pi\,\left(D + \rho_s - 4\,\epsilon_s - 4\,\rho_s\,
\frac{\Phi_{11}}{3\,\Psi_2 + 2\,\Phi_{11}}\right)\,{\cal 
T}_{2a}\,{T_{22}}^{(1)} 
%\nonumber \\
%&& 
+
\frac{\pi}{\varphi_1}\left(3\,\Psi_2 + 2\,\Phi_{11}\right)J_{2}^{(1)}, 
\label{eq:psi3_radang}
\end{eqnarray}
%}
%
where we have expanded the radial-temporal operators.
Under the hypothesis of radially infalling matter the equations simplify and 
it is possible to separate the angular dependence from the 
radial-temporal part. In order to get this decomposition we expand each relevant function
into a basis of spin weighted spherical harmonics as follows. The 
$\Psi_4^{(1)}$ function has spin weight $-2$,
$\Psi_3^{(1)}$ spin weight $-1$, and the density of matter is a scalar with spin weight zero. 
Expanding each function in 
the corresponding basis one gets
\begin{eqnarray}
{\Psi_4}^{(1)}&=&\sum_{l,m}\,P_4(t,r)\,{Y_{-2}}^{(l,m)}(\theta, \varphi),  \\
{\Psi_3}^{(1)}&=&\sum_{l,m}\,P_3(t,r)\,{Y_{-1}}^{(l,m)}(\theta, \varphi),  \\
\rho&=&\sum_{l,m}\,\rho(t,r)\,{Y_0}^{(l,m)}(\theta, \varphi).\label{eq:expansion}
\end{eqnarray}
%as ${T_{22}}^{(1)}=\rho(t,r)\,\left(k_\mu\,u^\mu\right)^2$, and ${J_{2}}^{(1)}=\rho(t,r)\,\left(k_\mu\,u^\mu\right)$.

These expansions are convenient because the angular operators of 
Eqs.~(\ref{eq:psi4_radang})-(\ref{eq:psi3_radang}) can be written in terms of 
the raising and lowering spin operators:
$\eth_s=-\left(\partial_\theta + i\,\csc\theta\,\partial_\varphi - s\,\cot\theta\right)$ 
and $\bar{\eth}_s=-\left(\partial_\theta - i\,\csc\theta\,\partial_\varphi + 
s\,\cot\theta\right)$  \cite{Newman66,Goldberg:1966uu}:
\begin{equation}
\delta + q\,\beta_s=-\frac{1}{r\,\sqrt{2}}\,\eth_{\frac{q}{2}}, \hspace{1cm}  \bar{\delta} + q\,\beta_s=-\frac{1}{r\,\sqrt{2}}\,
\bar{\eth}_{-\frac{q}{2}},
\end{equation}
from these last equations and Eq.~(\ref{def:ops}) with 
$\pi_s=\tau_s=\gamma_s=0$, $\alpha_s=-\beta_s$ and $\eta=-1$ one gets
\begin{eqnarray}
{\cal O}_{2a}&=&-(\delta - 4\,\beta_s)=\frac{1}{\sqrt{2}\,r}\,\eth_{-2}, 
\qquad {\cal O}_{3b}=(\overline{\delta} + 2\,\beta_s 
)=-\frac{1}{\sqrt{2}\,r}\,\bar{\eth}_{-1},  \nonumber \\
\overline{\delta} + 2\,\eta\,\alpha&=&\overline \delta + 2\,\beta_s=
-\frac{1}{\sqrt{2}\,r}\,\bar{\eth}_{-1}, \qquad 
{\cal T}_{2a}=\overline{\delta}=-\frac{1}{\sqrt{2}\,r}\,\bar{\eth}_{0},  
\nonumber \\
{\cal O}_{5b}&=&\delta  - 4\,\beta_s=-\frac{1}{\sqrt{2}\,r}\,\eth_{-2},  
\qquad
{\cal O}_{1b}=
-(\overline{\delta}+2\,\beta_s)=\frac{1}{\sqrt{2}\,r}\,\bar{\eth}_{ -1}. 
 \nonumber
\end{eqnarray}

When the $\eth$ operator
acts on the spin weighted spherical harmonic, it raises the spin weight:
\begin{equation}
\eth_s\,{Y_s}^{l,m}=\sqrt{\left(l-s\right)\,\left(l+s+1\right)}\,{Y_{s+1}}^{l,m}~, \label{op:eth_Y}
\end{equation}
and, when ${\bar \eth}$ acts on the spin weighted spherical harmonic, it lowers the spin weight:
\begin{equation}
{\bar \eth}_s\,{Y_s}^{l,m}=-\sqrt{\left(l+s\right)\,\left(l-s+1\right)}\,{Y_{s-1}}^{l,m}~. \label{op:beth_Y}
\end{equation}
For a detailed description of these harmonics and on how to use them to extract 
physical information carried by gravitational waves see for instance 
\cite{Ruiz:2007yx} and references therein. 
Notice that in terms of the harmonic coefficients for the density, we have that 
\begin{eqnarray}
[{\cal O}_{3b}{\cal T}_{2a}]\,T_{22}^{(1)}&=&\frac{1}{2\,r^2}\,\left(k_\mu\,u^\mu\right)^2\,\sum_{l,m}\,\rho_{l\,m}\,\bar{\eth}_{-1}\,\bar{\eth}_0
\,Y_0^{l,m}=\left(k_\mu\,u^\mu\right)^2\,\sum_{l,m}\,\rho_{l\,m}\,\frac{\sqrt{\left(l-1\right)\,l\,\left(l+1\right)\,\left(l+2\right)}}{2\,r^2}
\,\,Y_{-2}^{l,m}, \\
{\cal T}_{2a}\,{T_{22}}^{(1)} &=& \frac{1}{\sqrt{2}\,r}\,\bar{\eth}_0\,\rho\,\left(k_\mu\,u^\mu\right)^2=\left(k_\mu\,u^\mu\right)^2\,
\frac{1}{\sqrt{2}\,r}\,\sum_{l,m}\,\rho_{l\,m}\,\bar{\eth}_0\,Y_0^{l,m}=
\left(k_\mu\,u^\mu\right)^2\,\sum_{l,m}\,\frac{\sqrt{l\,\left(l+1\right)}}{\sqrt{2}\,r}\,\rho_{l\,m}\,Y_{-1}^{l,m},\\
J_{2}^{(1)}&=&-\bar{\delta}\,J_n^{(1)}=k_\mu\,u^\mu\,\frac{1}{\sqrt{2}\,r}\,\sum_{l,m}\,\rho_{l\,m}\,\bar{\eth}_0\,Y_0^{l,m}=
k_\mu\,u^\mu\,\sum_{l,m}\,\frac{\sqrt{l\,\left(l+1\right)}}{\sqrt{2}\,r}
\,\rho_{l\,m}\,Y_{-1}^{l,m}.
\end{eqnarray}
Substituting the expansions Eq. (\ref{eq:expansion}) into Eqs.~(\ref{eq:psi4_radang})-(\ref{eq:psi3_radang}), 
taking in consideration the eigenvalues of the angular operators 
and integrating over the solid angle, we can obtain an equation for each mode 
$(l, m)$ for $P_4$:

%
%\small{
\begin{eqnarray}
&&\left\{\left[\Delta + 
\mu_s\,
\chi
%\frac{3\,\Psi_2 - 2\Phi_{11}}{3\,\Psi_2 + 2\Phi_{11}}
\,
\left(5 + \frac{4\,\Phi_{11}\,\left(9\,\Psi_2 - 14\,
\Phi_{11}\right)}{\left(3\,\Psi_2 - 2\Phi_{11}
\right)^2}\right) \right]\,(D +\rho_s - 4\epsilon_s) -
% \frac{3\,\Psi_2 - 2\Phi_{11}}{3\,\Psi_2 + 2\Phi_{11}}
\chi
 \,\frac{\left(l-1\right)
 \,\left(l+2\right)}{2\,r^2} + 3\,\Psi _{2} - 2\,\Phi _{11}
\right\}\,P_4(t,r) \nonumber \\
&& -\frac{4\,\Phi _{11}}{3\,\Psi_2 + 2\Phi_{11}}\,
\left(\Delta - \frac{9\,\Psi_2 - 14\,\Phi_{11}}{3\,\Psi_2 - 2\Phi_{11}}\,
\mu_s\right)\,
\frac{\sqrt{\left(l-1\right)\,\left(l+2\right)}}{\sqrt{2}\,r}\,P_3(t,r)
= \nonumber \\&&-4\,{\pi}\,\chi \,
%\frac{3\,\Psi_2 - 2\Phi_{11}}{3\,\Psi_2 + 2\Phi_{11}}\,
\frac{\sqrt{\left(l-1\right)\,l\,\left(l+1\right)\,\left(l+2\right)}}{2\,r^2}\,
\rho(t,r)\,(u^\mu\,k_\mu)^2.\nonumber \\  
\end{eqnarray}
and for $P_3$:
\begin{eqnarray}
&&\left[\left(D + 2\,\rho_s - 4\,\epsilon_s - 4\,\rho_s\,
\frac{\Phi_{11}}{3\,\Psi_2 + 2\,\Phi_{11}}\right)\,\left(\Delta - 
4\,\mu_s\right) 
%+
%\left(\frac{3\,\Psi_2 + 2\,\Phi_{11}}{3\,\Psi_2 - 2\,\Phi_{11}}\right)\,
-\frac{1}{\chi}
\frac{\left(l-1\right)\,\left(l+2\right)}{2\,r^2} 
+ 6\,\Psi_2 + 4\,\Phi_{11}\right]\,P_3(t,r) \nonumber \\ 
&&- 4\,\frac{\Phi_{11}}{3\,\Psi_2 - 2\,\Phi_{11}}\left[D + \rho_s - 
4\,\epsilon_s + 
6\,\rho_s\,\frac{\Psi_2}{3\,\Psi_2 + 2\,\Phi_{11}}\right]\,
\frac{\sqrt{2\,\left(l-1\right)\,\left(l+2\right)}}{2\,r}\,P_4(t,r)  = \\
&& 4\,\hat{\pi}\,\left(D + 2\,\rho_s - 4\,\epsilon_s - 4\,\rho_s\,
\frac{\Phi_{11}}{3\,\Psi_2 + 2\,\Phi_{11}}\right)\,
\frac{\sqrt{l\,\left(l+1\right)}}{\sqrt{2}\,r}\,\rho(t,r)\,(u^\mu\,k_\mu)^2
+ \frac{\pi}{\varphi_1}\left(3\,\Psi_2 + 
2\,\Phi_{11}\right)\,\frac{\sqrt{l\,\left(l+1\right)}}{\sqrt{2}\,r}\,\rho(t,r)\,
(u^\mu\,k_\mu). \nonumber 
\end{eqnarray}
Replacing directional derivatives and spin coefficients for their 
explicit form in the specified coordinates we obtain an explicit system for the 
radial and temporal part of each mode 
of the perturbations $\Psi_4^{(1)}$ and $\Psi_3^{(1)}$. In section 
\ref{sec:numerical} we will write the system down explicitly but first we will 
focus on 
the matter 
that produces the perturbations.
%
%%%%%%%%%%%%%%%%%%%%%%%%%%%%%%%%%%%%%%%%%%%%%
\section{Matter Content}
\label{sec:MD}
%%%%%%%%%%%%%%%%%%%%%%%%%%%%%%%%%%%%%%%%%%%%%

The dynamics of the particles is described by the conservation of the number of particles
and the conservation equation for the stress energy tensor $T^{\mu\nu}$. The 
continuity equation holds because we are assuming conservation 
of the number of particles, the particles belong to the same species and are not 
created or annihilated 
\begin{equation}
 \nabla_{\mu} J^{\mu} = 0 \ ,
\label{eq:conservation}
 \end{equation}
where $J^{\mu} = \rho u^{\mu}$.
The conservation equation \eqref{eq:conservation} for radially infalling 
particles yields
\begin{equation}
 \frac{1}{\sqrt{-g}} \partial_{\mu} \left( \sqrt{-g}\rho u^{\mu} \right) = 0 \ , \quad \Rightarrow
 \quad \partial_{t} ( r^2 \rho u^{t}) + \partial_{r} ( r^2 \rho u^{r}) = 0 \ ,
\end{equation}
which can be rewritten, given the metric \eqref{eq:metric} as:
\begin{equation}
 \partial_{t} \rho + v^{r} \, \partial_{r} \rho + \frac{\rho v^{r}}{r (u^{r})^2} 
 \left[ \left(E-\frac{qQ}{r}  \right) \left(2E-\frac{qQ}{r} \right) -2 +\frac{3M}{r}-\frac{Q^2}{r^2} \right] = 0\ ,
 \label{eq:cont}
\end{equation}
where we have defined $v^{r} \equiv\frac{u^{r}}{u^{t}}$. 
Since this is an equation that involves only temporal and radial derivatives, 
each mode $\rho_{l,m}$ in the decomposition (\ref{eq:expansion}) obeys 
Eq.~(\ref{eq:cont}) once we can provide the velocity
of each particle. In order to get
this velocity, we will use the conservation of the stress energy tensor.
First, notice that the equations%
\begin{equation}
T^{\mu\nu}{}_{;\mu}=0 \ ,
\label{eq:tmunu}
\end{equation}
can be integrated once, and the components of the four velocity can 
be expressed in terms of the constants of motion using the symmetries of the 
space-time and the normalization on the four-velocity. This can be achieved by 
noticing that equations~(\ref{eq:tmunu}) can also be obtained using 
the Euler-Lagrange equations with the Lagrangian
\begin{equation}
 \mathcal{L}= \mu \left(\frac{1}{2} g_{\mu\nu} u^{\mu} u^{\nu} + q A_{\mu} u^{\mu} \right),
\label{eq:lagrangian}
 \end{equation}
where $A_{\mu}=-Q/r d t$ is the vector potential.
The Euler-Lagrange equations become 
\begin{equation}
 u^{\mu} \nabla_{\mu} u^{\nu} = q F^{\nu}{}_{\alpha}u^{\alpha},
\end{equation}
where $F_{\mu\nu} = A_{\mu;\nu}-A_{\nu;\mu}$. 

Since the Lagrangian Eq.~(\ref{eq:lagrangian}) is independent of time, we get a 
constant of motion 
\begin{equation}
 -\varepsilon \equiv \frac{\partial}{\partial u^{t}} \mathcal{L}  \ ,\quad  
E=\frac{\varepsilon}{\mu}, \label{eq:energy}
\end{equation}
and considering the four velocity as
$u^{\mu} = (u^{t},u^{r},0,0)$ we get from Eq.\eqref{eq:energy}:
\begin{equation}
 u^{t} = \frac{Er^{2}+(2Mr-Q^{2})u^{r} +q A_{t}r^{2}  }{r^2-2Mr+Q^{2}}.
 \label{eq:ut_up}
\end{equation}
The normalization of the four velocity $u^{\mu} u_{\mu}=-1$ together 
with Eq.\eqref{eq:ut_up}, allow us to compute the radial component of the 
velocity. After some algebraic steps we get
\begin{equation}
( u^{r} )^{2} = (E+{q}A_{t})^{2} - \left( 1-\frac{2M}{r}+\frac{Q^2}{r^2}\right).
\label{eq:ur_up}
\end{equation}
Eq. \eqref{eq:ur_up} can be rewritten in terms of an effective potential
\begin{equation}
 \left(\frac{dr}{d\tau}  \right)^{2} = E^2 - V_{{\rm eff}}\ ,
\end{equation}
where 
\begin{eqnarray}
 V_{{\rm eff}}&=&1-\frac{2 M}{r}-\frac{q^2 Q^2}{r^2}+\frac{2 q Q E }{r}
 +\frac{Q^2}{r^2}, \nonumber\\
&=&1 -\frac{2M}{r}\left(1-\frac{EqQ}{M} \right) +\frac{Q^2}{r^2}\left(1-q^2\right).
 \end{eqnarray}
Notice that taking $E=1$, an extremal particle ($q=1$) in an extreme black 
hole ($Q=M$) will be in equilibrium \cite{Townsend:1997ku}.
The previous decomposition allows us to study, with a 1D numerical code, any 
radially in-falling dust matter distribution and its gravitational reaction.

%%%%%%%%%%%%%%%%%%%
\section{Numerical implementation}
\label{sec:numerical}
%%%%%%%%%%%%%%%%%%%

According to the {\it Peeling theorem} \cite{Wald84}, the Weyl scalars behave as
\begin{equation}
\Psi_i \equiv \frac{1}{r^{5-i}}.
\end{equation}
As such, it probes convenient to perform the evolution of the quantities $r\,{\Psi_4}^{(1)}$, and
$r^2\,{\Psi_3}^{(1)}$, so that the evolved quantity maintains a constant 
amplitude during the evolution. The evolution equation for $R_4=r\,P_4$
is 
\begin{eqnarray}
&&\left\{(r^2+2\,M\,r- Q^2)\,\frac{\partial^2}{\partial t^2} - \left(r^2-2\,M\,r 
+ Q^2\right)\,
\frac{\partial^2}{\partial r^2} - 2\,\left(2\,M\,r - 
Q^2\right)\,\frac{\partial^2}{\partial t \partial r} + \right. \nonumber \\
&&  \left. - 
2\,\frac{\left(6\,M\,r^3 + \left(3\,M^2 - 10\,Q^2\right)r^2  - 5\,M\,Q^2\,r  - 
2\,Q^4\right)}{r\,(3\,M\,r - 4\,Q^2)}\,\frac{\partial}{\partial t} 
- 
2\,\frac{\left(6\,M\,r^3 - \left(3\,M^2 + 10\,Q^2\right)r^2  + 5\,M\,Q^2\,r  + 
2\,Q^4\right)}{r\,(3\,M\,r - 4\,Q^2)}\,\frac{\partial}{\partial r} + \right. 
\nonumber \\
&& \left. \left(l-1\right)\,\left(l+2\right)\,\frac{3\,M\,r - 4\,Q^2}{3\,M\,r - 
2\,Q^2} 
- \frac{6\,M\,\left(M\,r - 2\,Q^2\right)}{r\,\left(3\,M\,r - 
4\,Q^2\right)}\right\}\,R_4(t,r) \label{ec:S1_ff}  \nonumber \\ 
&&-2\,\frac{\sqrt{2\,\left(l-1\right)\,\left(l+2\right)}}{3\,M\,r - 
2\,Q^2}\,Q^2\,
\left(\frac{\partial}{\partial t} - \frac{\partial}{\partial r} + 
\frac{4\,Q^2}{r\,\left(3\,M\,r - 4\,Q^2\right)}\right)\,R_3(t,r)
= \nonumber \\ &&+ 
4\,{\pi}\,\sqrt{\left(l-1\right)\,l\,\left(l+1\right)\left(l+2\right)}\,
\frac{3\,M\,r - 4\,Q^2}{3\,M\,r - 2\,Q^2}\,r\,\rho(t,r)\,
(u_\mu\,k^\mu)^2,
\nonumber \\
\end{eqnarray}
and 
for $R_3=r^2\,P_3$:
\begin{eqnarray}
&&\left\{(r^2+2\,M\,r- Q^2)\,\frac{\partial^2}{\partial t^2} - 
\left(r^2-2\,M\,r + Q^2\right)\,
\frac{\partial^2}{\partial r^2} - 2\,\left(2\,M\,r - 
Q^2\right)\,\frac{\partial^2}{\partial t \partial r} + \right. \nonumber \\
&&  \left. - 2\,\frac{3\,M\,r^3 - Q^2\,r^2 + M\,Q^2\,r - Q^4}{\left(3\,M\,r - 
2\,Q^2\right)\,r}\,\frac{\partial}{\partial t}  
- 2\,\frac{\left(3\,r^2 - Q^2\right)\,\left(M\,r - Q^2\right)}{\left(3\,M\,r - 
2\,Q^2\right)\,r}\,
\frac{\partial}{\partial r} 
\right. \nonumber \\
&& \left. + \left(l-1\right)\,\left(l+2\right)\,\frac{3\,M\,r - 2\,Q^2}{3\,M\,r 
- 4\,Q^2} 
- \frac{\left(3\,M\,r^3 - 6\,\left(3\,M^2 + Q^2\right)\,r^2 + 23\,M\,Q^2\,r - 
6\,Q^4\right)}{\left(3\,M\,r - 2\,Q^2\right)\,r^2}\right\}  
\,R_3(t,r)  \label{ec:S2_f}  \nonumber \\ 
&&-\frac{\sqrt{2\,\left(l-1\right)\,\left(l+2\right)}}{3\,M\,r - 4\,Q^2}\,
Q^2\,\left[{\left(r^2 + 2\,M\,r - Q^2\right)}\,\frac{\partial}{\partial t} + 
{\left(r^2 - 2\,M\,r + Q^2\right)}\,\frac{\partial}{\partial r} 
%\right. \nonumber \\ && \left. 
- 2\,Q^2\,\frac{r^2 + M\,r - Q^2}{\left(3\,M\,r - 2\,Q^2\right)\,r}\right]
\,R_4(t,r) 
=  \nonumber \\
&&{\pi}\,\sqrt{2\,l\,\left(l+1\right)}\,r\,\left[2\,\left(u_\mu\,k^\mu\right)^2\
, \left(r^2 + 2\,M\,r - Q^2\right)\,
\frac{\partial\,\rho(t,r)}{\partial\,t}
+ \left(r^2 - 2\,M\,r + 
Q^2\right)\,\frac{\partial\,\left(\rho(t,r)\,\left(u_\mu\,k^\mu\right)^2\right)}
{\partial\,r} 
\right. \nonumber \\
&& \left. - u_\mu\,k^\mu\,\left(4\,u_\mu\,k^\mu\,\frac{2\,M\,r^3 - 
2\,\left(2\,M^2 + Q^2\right)\,r + 9\,M\,Q^2\,r - 4\,Q^4}{(3\,M\,r - 2\,Q^2)\,r}
- \sqrt{2}\,\frac{3\,M\,r - Q^2}{Q}\right) \rho(t,r)\right].
\end{eqnarray}
To obtain a first order system of equations of motion suitable for 
numerical integration,
we introduce the auxiliary functions
\begin{equation}
 \pi_a \equiv \frac{1}{\alpha^2} (\partial_t R_a -\beta^{r}\psi_a)\ , \quad 
\psi_a\equiv \partial_r R_a \ , \qquad {\rm with}\quad a=3,4,
\end{equation}
where we have dropped the mode number subscripts to
simplify the notation and used the lapse and shift vector for the 
metric Eq.~(\ref{eq:metric})
\begin{eqnarray}
 \alpha=\left(1+\frac{2M}{r}-\frac{Q^2}{r^2} \right)^{-1/2} \ , \qquad \beta_{r} =
\frac{2M}{r}-\frac{Q^2}{r^2}\ , \qquad \beta^{r}=\alpha^2\beta_{r} \ .
\end{eqnarray}

The following system of evolution
equations for the functions $\pi_4$, $\psi_4$, $R_4$ is obtained,
\begin{equation}
 \partial_{t} R_a = \alpha^2 \pi_a +\beta^r \psi_a,\\ \qquad {\rm with }\quad 
a=3,4, \label{fo1}
\end{equation}
\begin{equation}
 \partial_{t} \psi_a = \alpha^2 \partial_{r}\pi_a +\beta^{r}\partial_{r}\psi_a +\frac{2r(Mr-Q^2)}{(r^2+2Mr-Q^2)^2}(\pi_a-\psi_a) \ , \quad a=3,4, \label{fo2}
\end{equation}
\begin{eqnarray}
 \partial_t\pi_4 &=& \beta^r\partial_r\pi_4+\alpha^2\partial_r\psi_4 +  
 {\beta_r}(\pi_4\partial_r\alpha^2+\psi_4\partial_r\beta^r)+2C_{4t}(\alpha^2\pi_4+\beta^{r}\psi_4 )-2C_{4r}\psi_4-C_{4in}R_4 \nonumber \\
 &+&2C_{3}\left(\alpha^2(\pi_3-\psi_3)+ \frac{4Q^2}{r(3Mr-4Q^2)}R_3\right)+\frac{3Mr-4Q^2}{r(3Mr-2Q^2)}\sqrt{(\ell+2)(\ell+1)\ell(\ell-1)}T_2 \ , \label{fo3} 
 \end{eqnarray}
where the coefficients are
\begin{equation*}
 C_{4t} =\frac{3Mr^2(M+2r)-Q^2(10r^2+2Q^2+5Mr)}{r^3(3Mr-4Q^2)}\ , \qquad  C_{4r} 
=\frac{ 3Mr^2(M-2r)-Q^2(-10r^2+2Q^2+5Mr)}{r^3(3Mr-4Q^2)}\ ,
\end{equation*}
\begin{equation*}
C_{4in} = \frac{(3Mr-4Q^2)(l+2)(l-1)   }{r^2(3Mr-2Q^2)} - \frac{6M((Mr-2Q^2))}{r^3(3Mr-4Q^2) } \ , 
\qquad  C_{3}=Q^2 \frac{\sqrt{2(l+2)(l-1)}}{r^2(3Mr-2Q^2)}\ .
\end{equation*}
For the remaining functions $\pi_3$, $\psi_3$, $R_3$ 
\begin{eqnarray}
\partial_t\pi_3 &=& \beta^r\partial_r\pi_3+\alpha^2\partial_r\psi_3 +  
\beta_r(\pi_3\partial_r\alpha^2+\psi_3\partial_r\beta^r)+2C_{3t}
(\alpha^2\pi_3+\beta^r\psi_3) - 2 C_{3r}\psi_3 - C_{3in}R_3 \nonumber \\
 &&+\frac{\sqrt{2(l+2)(l-1)}}{(3Mr-4Q^2)} \left[
 Q^2(\pi_4+\psi_4)
 -\frac{2Q^2(r^2+Mr-Q^2)  
}{r^3(3Mr-2Q^2)}R_4\right] \nonumber \\
 &&+\frac{r}{2}\sqrt{2l(l+1)} \left( \frac{1}{\alpha^2} \frac{\partial}{\partial 
t}T_2 
+\left( \alpha^2-\beta_r\beta^r  \right) \frac{\partial}{\partial r}T_2 
\right. \nonumber \\
&&\left.-\frac{ ( 6M^2r-5MQ^2-9Mr^2+8Q^2r )  }{r^2(3Mr-2Q^2)}T_2+  \left(  
\frac{3Mr-2Q^2}{r^2Q\sqrt{2}}     \right)J_2\right),
\end{eqnarray}
where $T_2=\rho_{l,m}\,\left(u^t+u^r\right)^2$, 
$J_2=\rho_{l,m}\,\left(u^t+u^r\right)$
and the coefficients
\begin{equation*}
 C_{3t}= 
 \frac{6M^2r^2- Q^2(-6Mr+Q^2+r^2)}{r^3(3Mr-2Q^2)} \ ,
\qquad
 C_{3r} = \frac{6Mr^2(M-r)+Q^2(-6Mr+Q^2+5r^2 )}{r^3(3Mr-2Q^2)} \ ,
 \end{equation*}
\begin{equation*}
C_{3in} = \frac{(3Mr-2Q^2)(l+2)(l-1)}{r^2(3Mr-4Q^2)}- \frac{( 
3Mr^3-18M^2r^2-6Q^2r^2+23MQ^2r-6Q^4)}{r^4(3Mr-2Q^2)}. 
 \end{equation*}

%%%%%%%%%%%%%%%%%%%
\section{Results and discussion}
\label{sec:results}
%%%%%%%%%%%%%%%%%%%
We solve the equations for the gravitational-electromagnetic field by using 
the method of lines. 
Our numerical code evolves the first order variables Eq. (\ref{fo1})-(\ref{fo3}) with a third order Runge Kutta integrator
with a fourth order spatial stencil in a finite computational domain $r \in [r_{\rm min},r_{\rm max}]$. 
We also introduce a small sixth order dissipation to eliminate high frequency modes. 
As boundary conditions at the last grid point, we impose that all the
incoming waves as given by the characteristic fields vanish.  
We set $r_{\rm max}$ sufficiently far out in order to avoid any kind of contamination, typically $r_{\rm max} >
t_{evol}$ where $t_{evol}$ is the total evolution time. 
Since we are using horizon penetrating coordinates
$r_{\rm min}$ lies inside the event horizon. 
We solve the equation for the rest mass density and look for the
gravitational and electromagnetic responses. 
Then coupled waveforms are then extracted at a fixed radius $r=r_{\rm obs}$.
In what follows we consider for simplicity, a shell of matter described by a single spherical harmonic mode
\begin{equation}
 \rho(t,r) = \rho_{l,m} Y_{0}^{l,m},
\end{equation}
we consider the modes with $l = 2$ since these give the main contribution to 
the quadrupole gravitational
radiation. We have used as initial data for the radial distribution for the density a Gaussian of the form
\begin{equation}
 \rho_{\ell,m}(t=0,r) = \rho_{0} e^{-(r-r_{\rm cg})^2/2\sigma^2} \ 
,\label{eq:gauss_rho}
\end{equation}
with $\rho_{0} = 5\times10^{-3}$, $r_{\rm cg} = 10M$ and $\sigma = 0.5M$.  

\begin{figure}[h!]
\begin{center}
%\vspace{-1.7cm}
\includegraphics[width=0.7\textwidth]{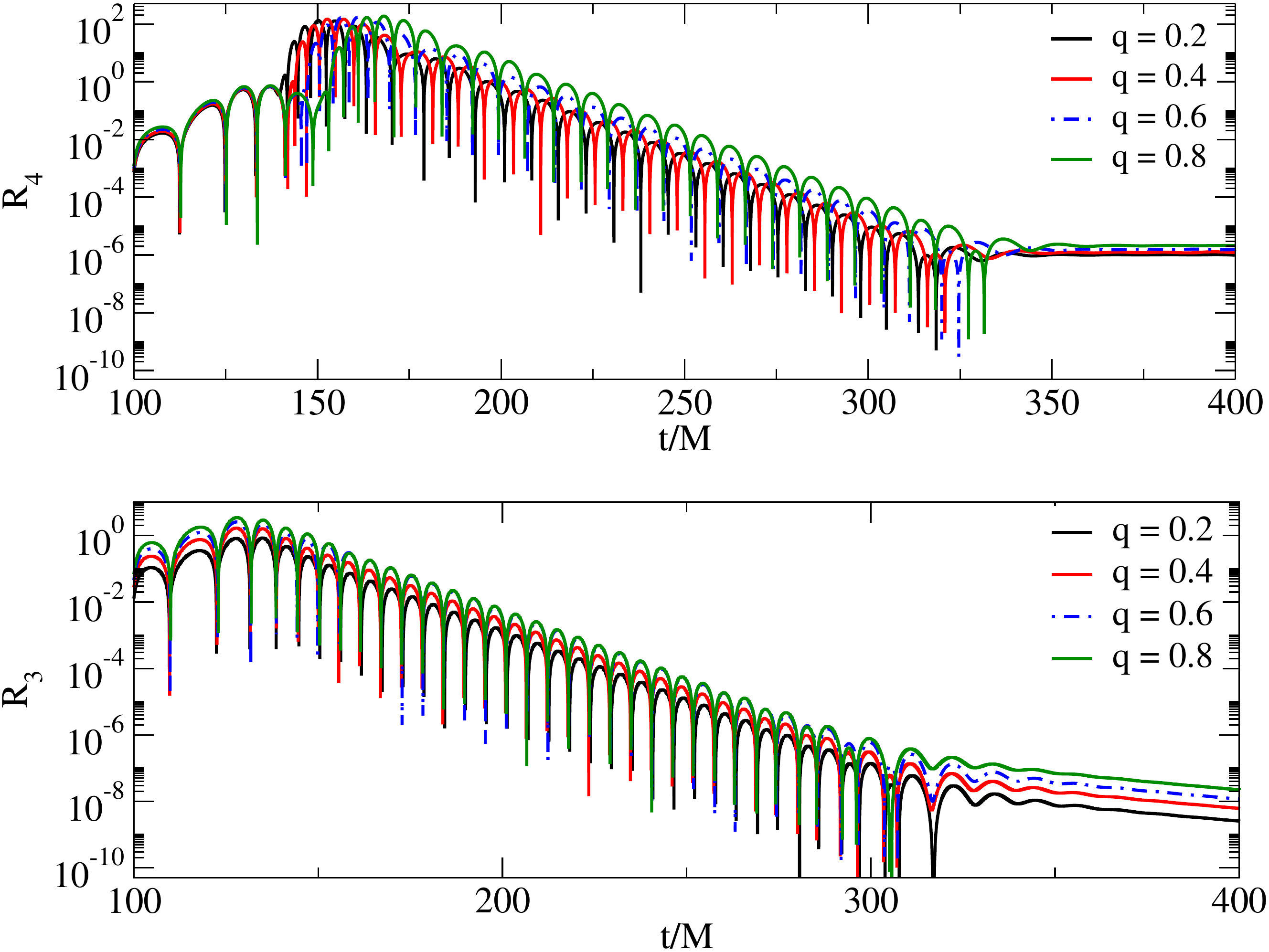}\\
\caption{The frequency of the waveforms $R_4$ (top) and $R_3$ 
(bottom) is independent of the value of $q$, it depends only on the charge and 
mass of the black hole and are the quasinormal modes. The difference in phase 
observed in $R_4$ is due to the difference in the infalling time of the charged 
particles.
}
\label{fig:gravandelect_vsq}
\end{center}
\end{figure}

Fig. \ref{fig:gravandelect_vsq} illustrates the waveform for a generic evolution. We
have used a Gaussian perturbation
centered at $r_cg$ as initial data at $t = 0$ for the radial distribution for the density.

The gravitational-electromagnetic functions $R_4$, $R_3$ are set to zero 
initially  as well as their derivatives. 
Our results indicate that this choice on the waveforms has negligible effects because the signals are ruled by the 
perturbation caused by the infalling matter.

\begin{figure}[h]
\begin{center}
%\vspace{-1.7cm}
\includegraphics[width=0.7\textwidth]{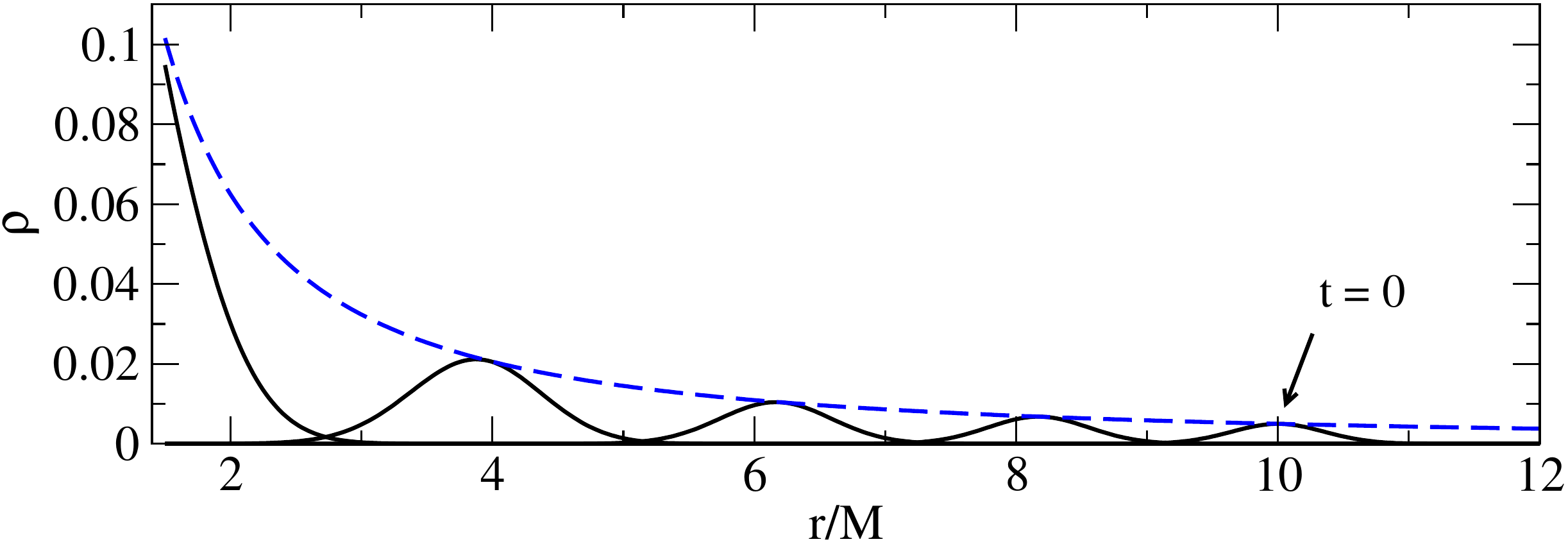}
\vspace{-0.3cm}\\
\caption{Snapshots of the evolution of the density given by 
\eqref{eq:cont} taken 
every $t=25M$. The dashed line represents the trajectory of the 
point with the highest density for the Gaussian initial 
data. The trajectory is modelled by~\eqref{eq:envelope}. For this plot we used 
$r_{cg}=10M$, 
$\sigma = 0.5M$, $q=0.2$ and $Q=0.9$.}
\label{fig:envelope_Q09_q02}
\end{center}
\end{figure}

Given an initial distribution of the infalling density $\rho_0$, centered at 
$r_{cg}$ Eq.
\eqref{eq:cont} can be integrated from $r_{cg}$ to $r$
to give the envelope of the density as
\begin{equation}
\rho =\rho_0 \frac{r_{cg}}{r} \left[ \frac{(q^2-1) Q^2+2r_{cg}(M-EqQ)+r_{cg}^2(E^2-1)}{(q^2-1) Q^2+2r(M-EqQ)+r^2(E^2-1)}  \right]^{1/2} \ .
\label{eq:envelope}
\end{equation}
In order to integrate Eq. \eqref{eq:cont} we use the fact that 
\begin{equation}
 \frac{d}{dt}\rho = \left({\partial_t}+v^r {\partial_r}\right)\rho,
\end{equation}
and $v^{r}=\frac{dr}{dt}$.
In Fig. \ref{fig:envelope_Q09_q02} we show the density profile at different 
times (i. e., every $t = 25M$) with the exact solution for the envelope  
Eq.~\eqref{eq:envelope}. This procedure allows us to prove the accuracy of our 
numerical code.

\begin{figure}[!ht]
\centering
%\vspace{-1.7cm}
\includegraphics[width=0.4\textwidth]{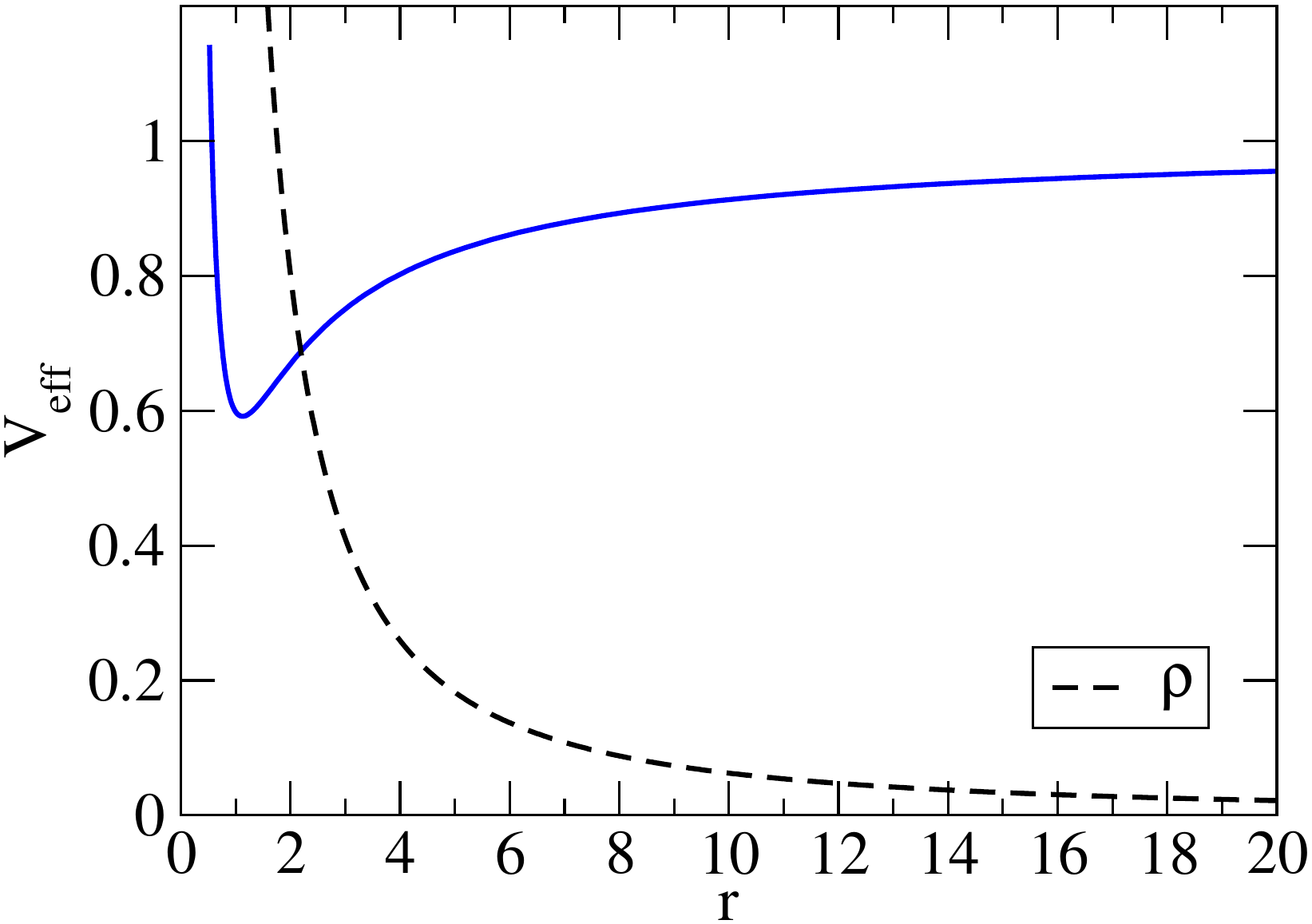} 
\vspace{0.5cm}\hspace{0.5cm}
\includegraphics[width=0.4\textwidth]{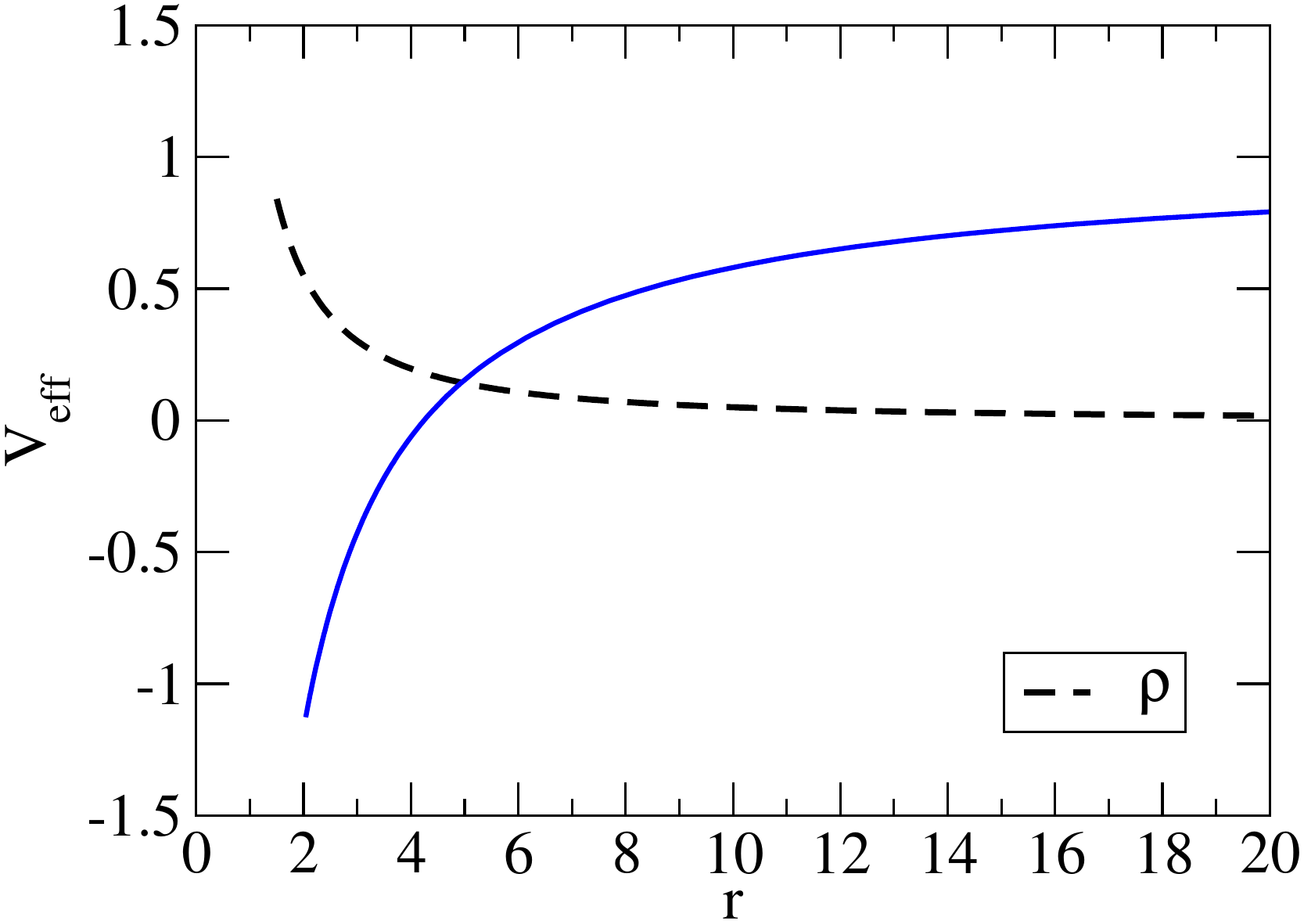} 
\vspace{0.5cm}\hspace{0.5cm}
\includegraphics[width=0.4\textwidth,height=5cm]{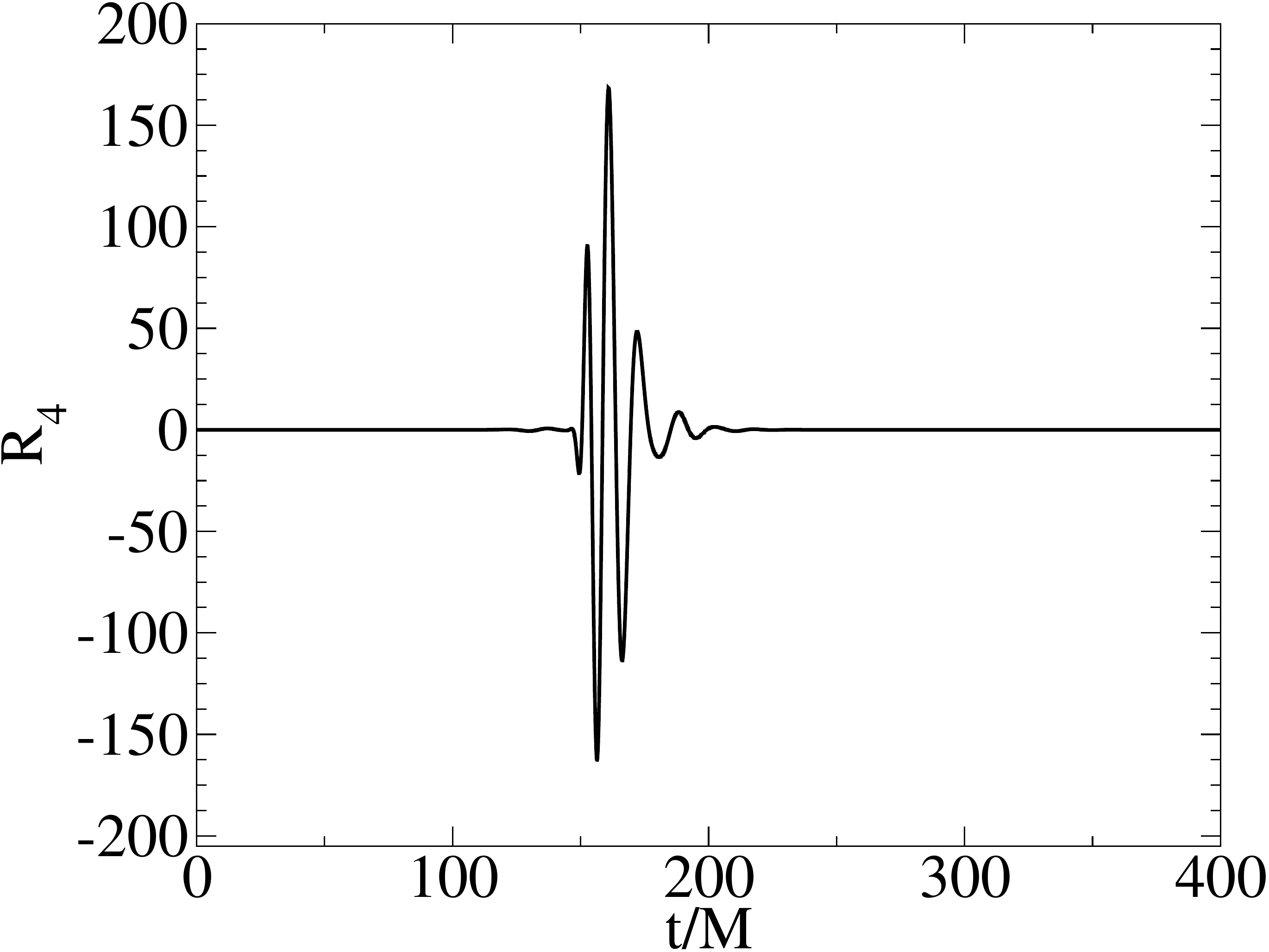} 
\vspace{0.3cm}\hspace{0.4cm}
\includegraphics[width=0.4\textwidth,height=4.95cm]
{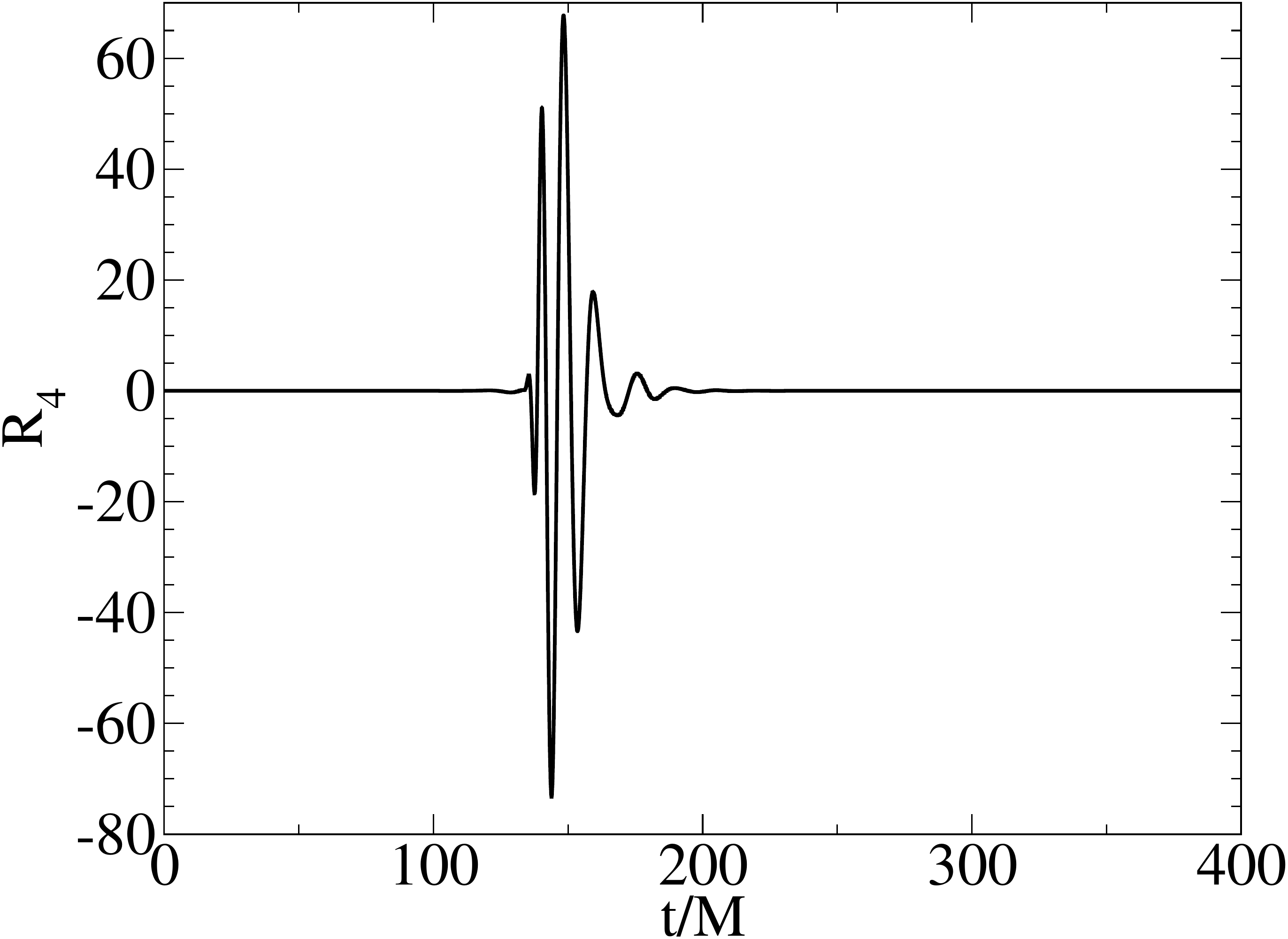} 
\caption{The potential $V_{\rm eff}$ is shown in the first row for two 
representative cases of the charge of 
the particles, $q=0.6$ and $q=-1.2$, the particles have $E=1$ and the charge of 
the BH is $Q=0.9$. The black line is the envelope of the density, \emph{i.e.} 
the trajectory of the point with higher density in the fluid. In the second row 
it 
is shown the waveform $R_4$ caused by the infalling on the particles. Although 
the dynamics of the particles is different, as can be infer from the potential, 
the waveforms are quiet similar in structure, the same happens for $R_3$.
}
\label{fig:rho_potQ09}
\end{figure}
%
%%%%%%%%%%%%%%%%%%%%%%%%%%%%%%

In Fig.~\ref{fig:rho_potQ09} we plot $V_{\rm eff}$, the 
density Eq.~(\ref{eq:envelope}) and the resulting waveforms 
$R_4$ for two representative values $q=0.6$, $q=-1.2$. For $q=0.6$ the 
potential has a minimum and the 
potential barrier lies outside the external horizon. For $q=-1.2$ the 
electromagnetic force contributes to the gravitational attraction and the 
particles fall faster onto the black hole. The effective potential thus 
presents very distinctive properties depending on the value of $q$. The 
situation with the gravitational-electromagnetic signals is different.
Although quantitatively the waveforms $R_4$ are different for $q=0.6$ and 
$q=-1.2$, they show qualitatively the same behavior, an initial burst, the 
ringdown phase and the tail despite the difference in the potential.
The main difference is in the amplitude of the wave and the time of response. 
The same follows for $R_3$. \\
%%%%%%%%%%%%%%%%%%%%%%%%%%%%%%55
\begin{figure}[h!]
\begin{center}
%\vspace{-1.7cm}
\includegraphics[width=0.7\textwidth]{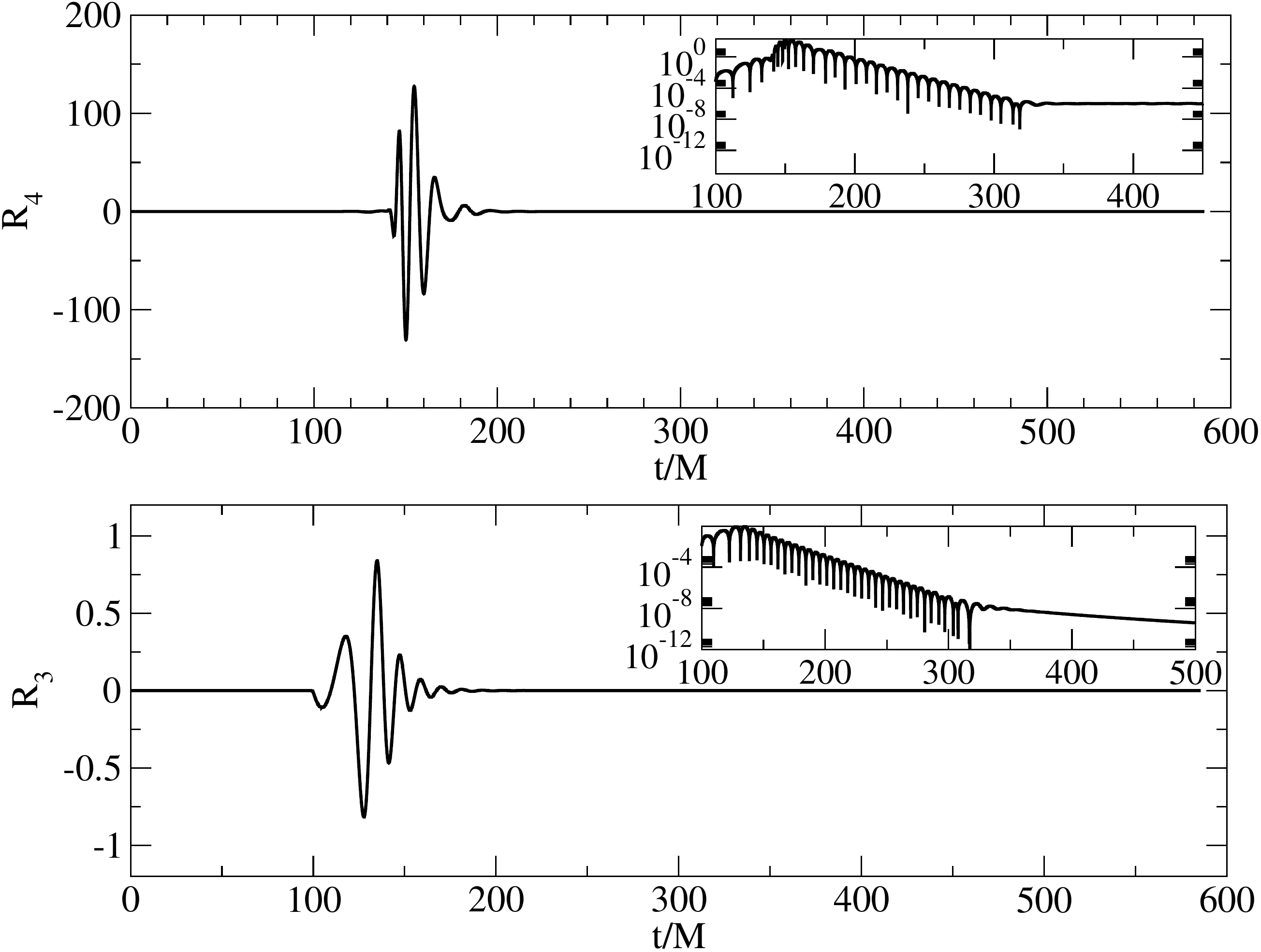}\\
\caption{Waveforms $R_4$ (top) and $R_3$ (bottom) emmited by a nearly extremal 
black hole with $Q=0.9$. The charge mass ratio of the particles has a moderate 
value $q=0.2$.  The inset displays 
the absolute values in a semi-logarithmic scale to show the rigndown and tail 
decay.
The observer in this example is placed in $r_{\rm obs}=100M$.
}
\label{fig:R1_Q09q02}
\end{center}
\end{figure}
In Fig. \ref{fig:R1_Q09q02} we show the signals $R_4$ and $R_3$ produced 
by the infalling matter Eq.~(\ref{eq:gauss_rho}) as measured by an observer located 
at  $r_{\rm obs}/M = 100$. For this plot, we use a value of $Q=0.9$ and 
$q=0.2$. For this relatively small value of $q$ the $R_4$ signal is stronger 
and displays the three characteristics phases of a gravitational wave emmited 
by a perturbed black hole: the initial burst, the quasinormal ringing and the 
power-law decay \cite{Kokkotas99a,Berti:2009kk,Konoplya:2011qq}. In the inset 
the signals are plotted in a 
semilogarithmic scale to improve the visualization of the two last phases. 
Fig. \ref{fig:gravandelect_vsq} shows the absolute value of $R_4$ and $R_3$ in 
a semilogarithmic scale for some values of the ratio $q$.  The frequency of 
$R_4$ does not change implying its is independence of $q$. The same result 
holds for $R_3$. The plot of $R_4$ however displays a shift in the time in 
which the signal is emitted. This result is consistent with the fact that the 
electromagnetic repulsion plays a role in the time of infall of the particles.  

%%%%%%%%%%%%%%%%%%%
\section{Conclusions}
\label{sec:conclusions}
%%%%%%%%%%%%%%%%%%%

The prime concern of this study has been the investigation, within linear 
perturbation theory, of waveforms of coupled electromagnetic and gravitational 
signals produced by point particles falling in the vicinity of the 
Reissner-N\"ordstrom black hole. 

The electromagnetic counterpart of gravitational waves caused by infalling 
charged particles into a neutral black hole was discussed in 
\cite{Degollado:2014dfa}.
In this work we extend that study to consider a charged black hole. Due to  
the electromagnetic interaction between the black hole 
and the particles it is expected that an electromagnetic wave will be emitted 
at the same time that the gravitational wave. 

As in \cite{Degollado:2014dfa}, the infalling of charged matter triggers 
gravitational and electromagnetic signals. In that work, 
we did not find a direct coupling between the frequencies of both types of 
waves. Since the spacetime was neutral. In this case, we 
allow the particles to interact the spacetime through the charge of the black 
hole and the charge of the particles, such coupling is reflected in the Weyl 
scalar $\Psi_3$.

We have shown that the signals, 
described by perturbations of the scalars $\Psi_3$ 
and $\Psi_4$ are coupled and the waveforms have the known phases: 
initial burst, quasinormal ringing and tail decay.

Since the particles consider here are charged, the electric field of the 
black hole affects their motion and they do not follow geodesics. However, the 
trajectories still have an analytic description and we
were able to follow the motion and provoke a gravitational and 
electromagnetic response, as the particles cross the horizon.

We have compared the electromagnetic and gravitational response when the black 
hole is not charged. The amplitude and shape of the waveforms are different 
from the neutral case. 
However, we estimate the quasinormal modes via a numerical fit of 
the signals and found that the frequencies do correspond to the 
quasinormal modes of the Reissner-Nordstrom black hole 
\cite{Chandrasekhar79, Andersson:1996xw, Onozawa:1995vu}.

The linear dependence of the electromagnetic waveforms with the charge mass 
ratio of the particles $q$ found in
\cite{Degollado:2014dfa} is also reproduced in 
this case. This result comes from the linear dependence of the perturbations 
with respect to $q$. 

In \cite{Degollado:2014dfa}, we had also found that the gravitational and 
electromagnetic energies were related by the square of $q$, which
pointed to ways of determining the energy of one type of wave if the other 
was measured. 
%{\bf Todavia pendiente...}
In the present
case, such relation does not hold anymore, as long as the energies due to the gravitational and 
to the electromagnetic response, are coupled, so that it is not possible to extract them independently. New 
forms of measuring the effects must be defined.

This last conclusion exemplifies the richness implicit in the Reissner-N\"ordstrom spacetime, and the dynamics
of the matter content. Several new relations and challenges are still open, as 
the different behavior of
the fields under the scope of different gauges. Some of these subjects will be dealt with in future work.

%%%%%%%%%%%%%%%%%%%%%%%%%%%%%%%%%%%%%%%%%%%%%%%
\section*{Acknowledgements}
%%%%%%%%%%%%%%%%%%%%%%%%%%%%%%%%%%%%%%%%%%%%%%%

We wish to thank Juan Carlos Hidalgo for comments on a previous version of our 
manuscript.
This work was partially supported by DGAPA-UNAM grant IN103514 and CONACYT  
grant No. 271904.
JCD acknowledges support from Instituto de Ciencias Fisicas, UNAM. 
CM acknowledges support from CONACYT-AEM grant No. 248411 and from 
PROSNI-UdG. The computations have been performed on
Miztli computer at DGTIC UNAM.

%%%%%%%%%%%%%%%%%%%%%%%%%%%%%%%%%%%%%%%%
%%%%%%%       References     %%%%%%%%%%%  
%%%%%%%%%%%%%%%%%%%%%%%%%%%%%%%%%%%%%%%%

\bibliography{referencias}

%%%%%%%%%%%%%%%
%%%   END   %%%
%%%%%%%%%%%%%%%

\end{document}